\documentclass{amsart}
\usepackage{amssymb}
\usepackage{tikz}
\usepackage{hyperref}
\usetikzlibrary{calc,decorations.pathmorphing,decorations.markings,
  decorations.pathreplacing,patterns,shapes}
\usepackage{color}

\allowdisplaybreaks[1]
\newcommand\xsup{\rm{x}}
\newcommand\tsup{\rm{t}}
\renewcommand\u{\uparrow}
\renewcommand\d{\downarrow}
\renewcommand\ss{\scriptstyle}
\newcommand\Rc{\check{R}}
\newcommand{\myid}{{\mathbf 1}}
\newcommand{\On}{{\rm O}(n)}
\newcommand{\ZN}{\mathbb{Z}_N}
\newcommand{\TL}{Temperley--Lieb}
\newcommand{\ie}{{\it i.e.}}
\newcommand{\cf}{{\it cf.\ }}
\newcommand{\wh}{\widehat}
\newcommand{\half}{\frac{1}{2}}

\newcommand{\ket}[1]{\left|#1\right>}
\newcommand{\aver}[1]{\left\langle #1 \right\rangle}
\newcommand{\Uq}[1]{U_q(#1)}
\newcommand{\thab}[2]{\Theta_{#1}{}^{#2}}
\newcommand{\thabhat}[2]{\widehat\Theta^{#1}{}_{#2}}
\newcommand\ad[2]{\mathop\mathrm{ad}\nolimits_{#1}\left( #2 \right)}
\newcommand{\ot}{\otimes}
\newcommand{\ra}{\rightarrow}
\newcommand{\cR}{\check{R}}

\newcommand{\secref}[1]{\S ~\ref{#1}}

\outer\def\slow#1{#1} 

\tikzset{edge/.style={very thick,draw=blue}}%
\tikzset{contour/.style={brown,dashed,postaction={decorate,decoration={markings,mark = at position #1 with {\arrow{>}}}}}}
\tikzset{contour/.default=0.5}

\def\loos{0.35}
\def\slt{0.2}
\pgfmathsetmacro{\ae}{atan(\slt)}
\pgfmathsetmacro{\aw}{\ae+180}
\pgfmathsetmacro{\an}{90-\ae}
\pgfmathsetmacro{\as}{\an+180}
\pgfmathsetmacro{\sltb}{sqrt(1-\slt*\slt)}
\pgfmathsetmacro{\lcrot}{45-atan(\slt/\sltb)*0.5}
\tikzset{distort/.style={cm={1,0,-\slt,\sltb,(0,0)}}}
\def\goI#1(#2,#3){
\pgfextra{
\pgfmathparse{#1+180}\global\let\oldangle=\currentangle\global\let\newangle=\pgfmathresult\global\let\currentangle=#1
\pgfmathparse{\oldx+#2}\global\let\newx=\pgfmathresult\xdef\oldx{#2}
\pgfmathparse{\oldy+#3}\global\let\newy=\pgfmathresult\xdef\oldy{#3}
}
.. controls ++(\oldangle:\loos) and ++(\newangle:\loos) .. ++(\newx,\newy)
}
\def\go#1{\expandafter\goI#1}
\def\startI#1(#2,#3){
\pgfextra{\global\let\currentangle=#1\xdef\oldx{#2}\xdef\oldy{#3}}}
\def\start#1{\expandafter\startI#1}
\def\north{\an(0,0.5)}
\def\south{\as(0,-0.5)}
\def\east{\ae(0.5,0)}
\def\west{\aw(-0.5,0)}

\tikzset{bgplaq/.style={fill=lightgray!20!white}}

\def\plaq(#1,#2){
\begin{scope}[shift={(#1,#2)}]
\draw[dotted] (-0.5,-0.5) rectangle ++(1,1); 
\end{scope}
}

\def\plaqz(#1,#2){
\begin{scope}[shift={(#1,#2)}]
\draw[bgplaq,dotted] (-0.5,-0.5) rectangle ++(1,1); 
\end{scope}
}

\def\plaqa(#1,#2){
\begin{scope}[shift={(#1,#2)}]
\draw[dotted,bgplaq] (-0.5,-0.5) rectangle ++(1,1);
\draw[edge] (0,-0.5) \start\north\go\east;
\draw[edge] (0,0.5) \start\south\go\west;
\end{scope}
}
\def\plaqb(#1,#2){
\begin{scope}[shift={(#1,#2)}]
\draw[dotted,bgplaq] (-0.5,-0.5) rectangle ++(1,1);
\draw[edge] (0,0.5) \start\south\go\east;
\draw[edge] (0,-0.5) \start\north\go\west;
\end{scope}
}

\tikzset{blob/.style={pos=0.5,circle,draw,fill=white,outer sep=0mm,inner sep=0.7mm}}
\def\bplaqe(#1,#2){
\begin{scope}[shift={(#1,#2)}]
\draw[dotted] (-0.5,0.5) -- (0.5,0.5) -- (-0.5,-0.5) -- cycle;
\end{scope}
}
\def\bplaqw(#1,#2){
\begin{scope}[shift={(#1,#2)}]
\draw[dotted] (0.5,-0.5) -- (0.5,0.5) -- (-0.5,-0.5) -- cycle;
\end{scope}
}
\def\bplaqn(#1,#2){
\begin{scope}[shift={(#1,#2)}]
\draw[dotted] (0.5,-0.5) -- (-0.5,0.5) -- (-0.5,-0.5) -- cycle;
\end{scope}
}
\def\bplaqs(#1,#2){
\begin{scope}[shift={(#1,#2)}]
\draw[dotted] (0.5,-0.5) -- (-0.5,0.5) -- (0.5,0.5) -- cycle;
\end{scope}
}
\def\bplaqez(#1,#2){
\begin{scope}[shift={(#1,#2)}]
\draw[dotted,bgplaq] (-0.5,0.5) -- (0.5,0.5) -- (-0.5,-0.5) -- cycle;
\end{scope}
}
\def\bplaqwz(#1,#2){
\begin{scope}[shift={(#1,#2)}]
\draw[dotted,bgplaq] (0.5,-0.5) -- (0.5,0.5) -- (-0.5,-0.5) -- cycle;
\end{scope}
}
\def\bplaqnz(#1,#2){
\begin{scope}[shift={(#1,#2)}]
\draw[dotted,bgplaq] (0.5,-0.5) -- (-0.5,0.5) -- (-0.5,-0.5) -- cycle;
\end{scope}
}
\def\bplaqsz(#1,#2){
\begin{scope}[shift={(#1,#2)}]
\draw[dotted,bgplaq] (0.5,-0.5) -- (-0.5,0.5) -- (0.5,0.5) -- cycle;
\end{scope}
}
\def\bplaqea(#1,#2){
\begin{scope}[shift={(#1,#2)}]
\draw[dotted,bgplaq] (-0.5,0.5) -- (0.5,0.5) -- (-0.5,-0.5) -- cycle;
\draw[edge] (0,0.5) \start\south\go\west;
\end{scope}
}
\def\bplaqeb(#1,#2){
\begin{scope}[shift={(#1,#2)}]
\draw[dotted,bgplaq] (-0.5,0.5) -- (0.5,0.5) -- (-0.5,-0.5) -- cycle;
\draw[edge] (0,0.5) \start\south\go\west
node[blob] {};
\end{scope}
}
\def\bplaqwa(#1,#2){
\begin{scope}[shift={(#1,#2)}]
\draw[dotted,bgplaq] (0.5,-0.5) -- (0.5,0.5) -- (-0.5,-0.5) -- cycle;
\draw[edge] (0,-0.5) \start\north\go\east;
\end{scope}
}
\def\bplaqwb(#1,#2){
\begin{scope}[shift={(#1,#2)}]
\draw[dotted,bgplaq] (0.5,-0.5) -- (0.5,0.5) -- (-0.5,-0.5) -- cycle;
\draw[edge] (0,-0.5) \start\north\go\east
node[blob] {};
\end{scope}
}
\def\bplaqna(#1,#2){
\begin{scope}[shift={(#1,#2)}]
\draw[dotted,bgplaq] (0.5,-0.5) -- (-0.5,0.5) -- (-0.5,-0.5) -- cycle;
\draw[edge] (0,-0.5) \start\north\go\west;
\end{scope}
}
\def\bplaqnb(#1,#2){
\begin{scope}[shift={(#1,#2)}]
\draw[dotted,bgplaq] (0.5,-0.5) -- (-0.5,0.5) -- (-0.5,-0.5) -- cycle;
\draw[edge] (0,-0.5) \start\north\go\west 
node[blob] {};
\end{scope}
}
\def\bplaqsa(#1,#2){
\begin{scope}[shift={(#1,#2)}]
\draw[dotted,bgplaq] (0.5,-0.5) -- (-0.5,0.5) -- (0.5,0.5) -- cycle;
\draw[edge] (0,0.5) \start\south\go\east;
\end{scope}
}
\def\bplaqsb(#1,#2){
\begin{scope}[shift={(#1,#2)}]
\draw[dotted,bgplaq] (0.5,-0.5) -- (-0.5,0.5) -- (0.5,0.5) -- cycle;
\draw[edge] (0,0.5) \start\south\go\east
node[blob] {}; 
\end{scope}
}

\def\plaqff(#1,#2){
\begin{scope}[shift={(#1,#2)}]
\draw[dotted,bgplaq] (-0.5,-0.5) rectangle ++(1,1);
\draw[edge] (0,-0.5) \start\north\go\east;
\end{scope}
}
\def\plaqf(#1,#2){
\begin{scope}[shift={(#1,#2)}]
\draw[dotted,bgplaq] (-0.5,-0.5) rectangle ++(1,1);
\draw[edge] (0,0.5) \start\south\go\west;
\end{scope}
}
\def\plaqd(#1,#2){
\begin{scope}[shift={(#1,#2)}]
\draw[dotted,bgplaq] (-0.5,-0.5) rectangle ++(1,1);
\draw[edge] (0,-0.5) \start\north\go\west;
\end{scope}
}
\def\plaqdd(#1,#2){
\begin{scope}[shift={(#1,#2)}]
\draw[dotted,bgplaq] (-0.5,-0.5) rectangle ++(1,1);
\draw[edge] (0,0.5) \start\south\go\east;
\end{scope}
}
\def\plaqc(#1,#2){
\begin{scope}[shift={(#1,#2)}]
\draw[dotted,bgplaq] (-0.5,-0.5) rectangle ++(1,1);
\draw[edge] (0,-0.5) \start\north\go\north;
\end{scope}
}
\def\plaqcc(#1,#2){
\begin{scope}[shift={(#1,#2)}]
\draw[dotted,bgplaq] (-0.5,-0.5) rectangle ++(1,1);
\draw[edge] (-0.5,0) \start\east\go\east;
\end{scope}
}
\def\plaqg(#1,#2){
\begin{scope}[shift={(#1,#2)}]
\draw[dotted,bgplaq] (-0.5,-0.5) rectangle ++(1,1);
\end{scope}
}

\pgfdeclaremetadecoration{prewavy}{zigzag}{
  \state{zigzag}[width=\pgfmetadecorationsegmentamplitude*\pgfmetadecoratedpathlength - 0.5*\pgfmetadecorationsegmentlength ,
                  next state=straight] {
        \decoration{zigzag}
      }
    \state{straight}[width=\pgfmetadecorationsegmentlength,
                   next state=final] {
        \decoration{curveto}
    }
    \state{final}{
        \decoration{zigzag}
        \beforedecoration{\pgfpathmoveto{\pgfpointmetadecoratedpathfirst}}
    }
}

\tikzset{
  wavy/.style={
    decorate,
    decoration={
      prewavy,
      meta-amplitude=#1,
      meta-segment length=0.3cm,
      amplitude=1.5pt, 
      segment length=6pt 
},
    postaction={decorate,ultra thick,decoration={markings,mark = at position #1 with {\arrow{>}}}}        
  },
  wavy/.default=0.5
}
\tikzset{oper/.style={rectangle,fill,inner sep=2.5pt}}
\tikzset{arr/.style={postaction={decorate,thick,decoration={markings,mark = at position #1 with {\arrow{>}}}}}}

\title{Discrete holomorphicity and quantized affine algebras}

 \author{Y.~Ikhlef}
 \author{R.~Weston}
 \author{M.~Wheeler}
 \author{P.~Zinn-Justin}

\address{Y.~Ikhlef, M.~Wheeler and P.~Zinn-Justin,
UPMC Univ Paris 6, CNRS UMR 7589, LPTHE,
75252 Paris Cedex, France.}

\address{R.~Weston, Department of Mathematics, Heriot-Watt University,
Edinburgh EH14 4AS, United Kingdom.}

\begin{document}

\begin{abstract}
We consider non-local currents in the context of quantized affine algebras, following the construction introduced by Bernard and Felder. In the case of $U_q(A_1^{(1)})$ and $U_q(A_2^{(2)})$, these currents can be identified with configurations in the six-vertex and Izergin--Korepin nineteen-vertex models. Mapping these to their corresponding Temperley--Lieb loop models, we directly identify non-local currents with discretely holomorphic loop observables. In particular, we show that the bulk discrete holomorphicity relation and its recently derived boundary analogue are equivalent to conservation laws for non-local currents.
\end{abstract}

\maketitle

\section{Introduction}
\label{sec:intro}
Discretely holomorphic observables are correlations functions in a two-dimensional
lattice model which satisfy a discrete version of the Cauchy--Riemann (CR) equations.
In the context of the Ising model, lattice fermions with this type of property
were first described in~\cite{DotsenkoP88}.
More recently, the discrete CR equations were used by Smirnov
as a basic tool to study rigorously the scaling properties of Ising
interfaces~\cite{Smir07}. They were then exploited in the Probability literature, to
obtain mathematical proofs of several Coulomb-gas results. For instance, still in the Ising model,
this approach yielded the scaling limit of domain walls and Fortuin-Kasteleyn cluster
boundaries~\cite{ChelkakS09,HonglerK11}, and the spin and energy correlation
functions~\cite{ChelkakHI12}.
For self-avoiding walks,
it provided a rigorous way to determine the bulk~\cite{DuminilS10} and
boundary~\cite{BeatondGG11} connectivity constants, and it has also proved very useful
for numerical purposes~\cite{BeatonGJ12}.
In~\cite{HonglerKZ12}, discretely holomorphic observables
for the Ising model have been related explicitly to the transfer-matrix formalism.

In the meantime, some discretely holomorphic observables have been identified
in other 2D lattice models, including the $\ZN$ clock model~\cite{RajC07} and the
dense~\cite{RivaC06} and dilute~\cite{IkhlefC09} {\TL} (TL) loop models.
These observables are essentially non-local, either because they include
disorder operators (in spin models), or because they are defined in terms of an
open path attached to the boundary (in loop models).
In all the known examples, it was observed that the discrete holomorphicity
condition is satisfied precisely when the Boltzmann weights are such that the model
is integrable~\cite{RajC07,IkhlefC09}. Recently, this statement was
also extended to the boundary Boltzmann weights~\cite{Ikhlef12,deGierLR12}.
These observed relations between the notions of discrete holomorphicity and
integrability have been explored further recently~\cite{AlamBatchelor12}, but they still
call for a more systematic understanding: this is the object of the present work.

An obvious starting point for this
is to try and construct discrete holomorphic observables from the underlying
symmetries of a the lattice model.
This idea is very reminiscent of the construction, proposed by Bernard and Felder~\cite{BernardFelder91},
of non-local conserved currents $\psi(z)$ in lattice models possessing
a quantum group symmetry. Indeed, in that context, $\psi(z)$ is non-local
because it includes a path connecting $z$ to a reference point, in a similar way
to disorder operators in spin systems. Moreover, the conservation equation 
for the current is a linear
relation between the values of $\psi$ at the points adjacent to a given vertex,
like the discrete holomorphicity condition. These resemblances~\cite{FendleyTalk12}
between discrete holomorphic observables and conserved currents can actually be made
more precise.

The present paper is based on a simple observation: in the case of an
affine quantum group symmetry, the conservation equation of currents
can be written as a discrete holomorphicity condition on the rhombic
lattice with opening angle $\alpha$, provided an
appropriate relation between $\alpha$ and the spectral parameter is introduced.
We thus consider
two simple loop models where discretely holomorphic observables -- which we shall call
for short {\it loop observables} -- are known:
the dense and dilute {\TL} models on the square lattice. 
Using the mapping of these loop models onto vertex models
possessing $\Uq{A_1^{(1)}}$ and $\Uq{A_2^{(2)}}$ symmetry respectively,
we construct the conserved currents, and
show that they map to the loop observables identified previously.
This analysis is extended to boundary observables in the case of general
diagonal integrable boundary conditions.

This point of view explains the somewhat mysterious observation
of~\cite{RajC07,IkhlefC09}
that discrete holomorphicity somehow ``linearizes'' the Yang--Baxter
equation by providing us with a linear equation for integrable
Boltzmann weights. Indeed, from our point of view, this linearization
procedure is nothing but Jimbo's interpretation~\cite{Jimbo86} of the $R$-matrix of an
integrable model as a representation of the universal $R$-matrix of a quantized affine algebra,
which by definition satisfies such a linear relation, as will be explained
below.

The plan of the paper is as follows. In \secref{sec:vertex}, we review the Bernard--Felder
construction of conserved currents introduced in~\cite{BernardFelder91} and expose a general identity 
for the adjoint action in this context.
\secref{sec:vertex-loop} reviews the correspondence between the six- and nineteen-vertex
models and the dense and dilute Temperley--Lieb loop models respectively. In particular, we show that 
the integrable weights can be obtained by solving the intertwining relations, which are linear equations
in the Boltzmann weights assigned to loop model tiles. In \secref{sec:DH-bulk}, we use the mapping between
vertex and loop models to express the currents as loop observables, satisfying the discrete Cauchy--Riemann
equations. \secref{sec:vertex-bound} extends the work of Bernard--Felder~\cite{BernardFelder91} to systems
with a boundary, and focuses on the interpretation of current conservation at the boundary.
\secref{sec:loop-bound} introduces boundary tiles into the dense and dilute \TL\ loop models.
Integrable weights for these tiles are obtained by solving linear equations which are the boundary
analogue of the intertwining relations.  In \secref{sec:DH-bound}, we express our currents
(which satisfy current conservation at the boundary) as loop observables obeying boundary
discrete Cauchy--Riemann equations. In \secref{sec:continuum}, we use the Coulomb-gas approach
to present the continuum limit of the loop observables. We conclude in \secref{sec:conclusion}.

\section{Vertex models, currents and quantized affine algebras}
\label{sec:vertex}
\subsection{Vertex models}
\label{ssec:lattice}
In this section, we recall how vertex models in statistical mechanics can be defined in terms of Boltzmann weights which are given as intertwiners of representations of quantized affine algebras. More specifically, we will consider the case in which we have a quantized affine algebra $U$ with a spectral parameter dependent module $V_z$. We denote the associated representation as $(\pi_z, V_z)$ where $\pi_z:U\rightarrow \hbox{End}(V_z)$ and additionally use the notation $\pi_{z_1,\dots,z_M}= \pi_{z_1}\otimes \cdots 
\otimes \pi_{z_m}$ for tensor products of such representations. 

Assuming that  $V_{z}\ot V_{w}$ is generically irreducible,
the $R$-matrix $R(z/w): V_{z}\ot V_{w}\rightarrow V_{z}\ot V_{w}$ that defines the vertex model is 
given as the solution (unique up to overall normalization) of the linear relation
\begin{equation}
R(z/w) \, \pi_{z,w}(\Delta(X)) 
=
\pi_{z,w} (\Delta'(X)) \, R(z/w)
\label{eq:Rcomm}
\end{equation} for all  $X\in U$; where $\Delta : U \rightarrow U \otimes U$ is the coproduct, $\Delta(X) = \sum X_1 \otimes X_2$, and $\Delta'$ the coproduct with the order of the tensor product reversed, $\Delta'(X) = \sum X_2 \otimes X_1$. 
We represent the $R$-matrix pictorially by
\begin{equation*}
R(z/w)=\begin{tikzpicture}[baseline=-3pt,scale=0.75]
\draw[arr=0.25] (0,-1) node[below] {$w$} -- (0,1);
\draw[arr=0.25] (-1,0) node[left] {$z$} -- (1,0);  
\end{tikzpicture}
\end{equation*}
The arrows drawn on the lines are purely to indicate ``time flow'': reading an equation from right to left corresponds to reading along a line in the direction of the arrows. 

Let us define the multiple coproduct $\Delta^{(L)}:U\ra U^{\ot(L)}$ by $\Delta^{(L+1)}=(\Delta\ot 1) \Delta^{(L)}$ and 
$\Delta^{(2)}=\Delta$. The monodromy matrix 
$
T^{(L)}(z;w_1,\dots,w_L): V_{z}\ot V_{w_1} \ot \cdots \ot V_{w_L}\ra 
V_{z} \ot V_{w_1} \ot \cdots \ot V_{w_L} 
$ 
is defined as
\[ 
T^{(L)}(z;w_1,\dots,w_L)= R_{0L}(z/w_L) \dots  R_{01}(z/w_1)\,,
\]
where the subscripts on $R$-matrices indicate the evaluation modules in which they act, \ie, $0 \leftrightarrow V_z$, $1,\dots,L \leftrightarrow V_{w_1},\dots,V_{w_L}$. Its graphical representation is 
\begin{equation*}
T^{(L)}(z;w_1,\dots,w_L)=\begin{tikzpicture}[baseline=-3pt,scale=0.75]
 \foreach\x in {1,2}
\draw[arr=0.25] (\x,-1) node[below] {$w_{\x}$} -- (\x,1);
 \foreach\x in {3}
\draw[arr=0.25] (\x,-1) node[below] {$\cdots$} -- (\x,1);
  \foreach\x in {4,...,8}
\draw[arr=0.25] (\x,-1) -- (\x,1);
 \foreach\x in {9}
\draw[arr=0.25] (\x,-1)  node[below] {$w_{L}$} -- (\x,1);
\draw[arr=0.05] (0,0) node[left] {$z$}  -- (10,0);
\end{tikzpicture}
\end{equation*}
Until specified otherwise, all lines will be oriented up/right in what follows,
so that we shall omit arrows on pictures.

By vertically concatenating $M$ monodromy matrices $T^{(L)}(z_i;w_1,\dots,w_L)$, $1\leq i \leq M$, one obtains a rectangular lattice $\Omega$ of width $L$ and height $M$. Each horizontal (resp.\ vertical) line of this lattice is oriented from left to right (resp.\ bottom to top), and denotes the vector space $V_{z_i}$ (resp.\ $V_{w_j}$).  

To simplify the discussion, from now on we will work on a horizontally and vertically homogeneous lattice, by assuming that all horizontal (resp.\ vertical) lines carry the same evaluation parameter $z_i = z$ (resp.\ $w_j = w$). We will be interested in operators that act on the vector spaces encoded by the lattice, which graphically corresponds to the insertion of a ``node'' at an edge. The edges of the lattice will be denoted by coordinates pairs $(x \pm \half,t)$ or $(x,t \pm \half)$, where $(x,t)$ is a vertex of $\Omega$, and $x$ (resp.\ $t$) is the horizontal (resp.\ vertical) coordinate.

Typically, one is interested in the case where the left/right boundaries of this lattice are fixed in some way. In such a case, each row of the lattice becomes an operator in $V_{w_1} \otimes \cdots \otimes V_{w_L}:= V_1 \otimes \cdots \otimes V_L$ that we will rather loosely call the ``one-row transfer matrix'' $\mathbf{T}$. We will use this construction below, despite the fact that for now we do not discuss the boundary conditions 
explicitly.

\subsection{Hopf algebras and graphical relations}
\label{ssec:hopf}
Following Bernard and Felder~\cite{BernardFelder91}, we consider a set of elements $\{J_a, \thab{a}{b}, \thabhat{a}{b}\}$, $a,b=1,\ldots,n$, of a Hopf algebra $U$. The elements $\thab{a}{b}$ and $\thabhat{a}{b}$ are assumed to be inverses of each other:
\begin{equation} 
\thab{a}{b} \thabhat{c}{b}= \delta_{a,c}
\quad
\hbox{and}
\quad
\thabhat{b}{a} \thab{b}{c}= \delta_{a,c}
\label{eq:inversion}
\end{equation}
(where here and subsequently repeated indices are summed over) while the coproduct $\Delta$ and antipode $S$ of all elements have the form: 
\begin{subequations}
\begin{align}
\label{eq:coprJ}
\Delta(J_a) &= J_a \otimes 1 + \Theta_a{}^b\otimes J_b
&
S(J_a)&=-\widehat\Theta^b{}_a J_b
\\
\label{eq:coprth}
\Delta(\Theta_a{}^c)&=\Theta_a{}^b\otimes \Theta_b{}^c
&
S(\Theta_a{}^b)&=\widehat\Theta^b{}_a
\\
\label{eq:coprhth}
\Delta(\widehat\Theta^a{}_c)&=\widehat\Theta^a{}_b\otimes \widehat\Theta^b{}_c
&
S(\widehat\Theta^a{}_b)&=\Theta_b{}^a.
\end{align}
It is also useful to define  $\wh J_a:=-\thabhat{b}{a} J_b$, which has the coproduct and antipode
\begin{align}
\label{eq:coprhJ} 
\Delta(\wh J_a) &=\wh J_b \otimes \thabhat{b}{a} + 1 \otimes \wh J_a
&
S(\wh J_a) &= \thabhat{c}{b} J_c \thab{a}{b}.
\end{align}
\end{subequations}
Given two elements in $\{J_a\}$,
say $J_1$ and $J_2$, $J_1+J_2$ can be trivially added to the set $\{J_a\}$ in such a way that (\ref{eq:coprJ}--\ref{eq:coprhJ}) still hold. It is a little
less obvious that the same is true of $J_1 J_2$, with appropriately defined sets
$\{\thab{a}{b}\}$ and $\{\thabhat{a}{b}\}$. Therefore we only need to specify a set $\{J_a\}$
that generates $U$ as a unital algebra.

The $R$-matrix, $R: U\otimes U \rightarrow U \otimes U$, switches the order of tensor products in the coproduct: namely, $R \Delta(X_a) = \Delta'(X_a) R$ for all $X_a \in U$. Applying this to the coproducts in (\ref{eq:coprJ}--\ref{eq:coprhJ}) gives, respectively
\begin{subequations}
\begin{align}
\label{eq:RcoprJ}
R (J_a \otimes 1 + \Theta_a{}^b\otimes J_b)
&=
(1 \otimes J_a  + J_b \otimes \Theta_a{}^b) R
\\
\label{eq:Rcoprth}
R (\Theta_a{}^b\otimes \Theta_b{}^c)
&=
(\Theta_b{}^c \otimes \Theta_a{}^b) R 
\\
\label{eq:Rcoprhth}
R (\widehat\Theta^a{}_b\otimes \widehat\Theta^b{}_c)
&=
(\widehat\Theta^b{}_c \otimes \widehat\Theta^a{}_b) R
\\
\label{eq:RcoprhJ} 
R (\wh J_b \otimes \thabhat{b}{a} + 1 \otimes \wh J_a)
&=
(\thabhat{b}{a} \otimes \wh J_b  + \wh J_a \otimes 1) R\,.
\end{align}
\end{subequations}
Suppose now that we have a representation $(\pi,V)$ of the Hopf algebra $U$. In the spirit of~\cite{BernardFelder91}, we can represent $\pi(J_a)$, $\pi(\thab{a}{b})$ and $\pi(\thabhat{a}{b})$ by the following pictures (from now on we always discuss representations of $U$, and so suppress the appearance of the $\pi$):
\begin{equation*}
J_a=\begin{tikzpicture}[baseline=-3pt,scale=0.75]
\draw(1,1) -- (1,-1);
\draw[wavy]
(0,0) node[below] {$a$} -- (1,0) node[oper] {} ;
\end{tikzpicture}
\, ,
\quad\quad
\thab{a}{b}=\begin{tikzpicture}[baseline=-3pt,scale=0.75]
\draw (1,1) -- (1,-1);
\draw[wavy]
 (0,0) node[below] {$a$}-- (2,0) node[below] {$b$}  ;
\end{tikzpicture}\,,
\quad\quad
\thabhat{a}{b}=\begin{tikzpicture}[baseline=-3pt,scale=0.75]
\draw (1,1) -- (1,-1);
\draw[wavy]
 (2,0) node[below] {$b$} --(0,0) node[below] {$a$} ;
\end{tikzpicture}
\end{equation*}
where the vertical line denotes the vector space $V$ with an upward arrow that we have suppressed, and subscripts (resp.\ superscripts) correspond to incoming 
(resp.\ outgoing) arrows. The operator $\wh J_a$ has the graphical
representation 
\begin{equation}
\wh J_a=-\ \begin{tikzpicture} [baseline=-3pt,scale=0.75]
\draw (0,-1.1) -- (0,0.5); \draw[wavy=0.4] (1,0) node[right] {$a$} -- (-1,0) -- (-1,-0.5) -- (0,-0.5) node[oper] {}; 
\end{tikzpicture}
\quad
\hbox{that we simplify to} 
\quad
\wh J_a=
\begin{tikzpicture} [baseline=-3pt,scale=0.75]
\draw (0,-0.7) -- (0,0.7); \draw[wavy=0.4] (1,0) node[right] {$a$} -- (0,0) node[oper] {}; 
\end{tikzpicture}\label{eq:jdual}
\end{equation}
so that a connection of a wavy line to a solid line
from the right (compared to the direction of time) corresponds to the
insertion of the operator $\wh J_a$. 

Using these notations, the equations listed have natural graphical meanings. Relation ~\eqref{eq:inversion} is expressed graphically by
\begin{equation}
\begin{tikzpicture} [baseline=-3pt,scale=0.75]
\draw (0,-0.75) -- (0,0.75);
\draw[wavy=0.35] (-1,0.3) -- (1,0.3) -- (1,0);
\draw[wavy=0.4] (1,0) -- (1,-0.3) -- (-1,-0.3);
\end{tikzpicture}
=
\begin{tikzpicture} [baseline=-3pt,scale=0.75]
\draw[wavy] (-1,0.3) -- (-0.5,0.3) to[out=0,in=0] (-0.5,-0.3) -- (-1,-0.3);
\draw (0,-0.75) -- (0,0.75);
\end{tikzpicture}
\quad
\hbox{and}
\quad
\begin{tikzpicture} [baseline=-3pt,scale=0.75,xscale=-1]
\draw (0,-0.75) -- (0,0.75);
\draw[wavy=0.35] (-1,0.3) -- (1,0.3) -- (1,0);
\draw[wavy=0.4] (1,0) -- (1,-0.3) -- (-1,-0.3);
\end{tikzpicture}
=
\begin{tikzpicture} [baseline=-3pt,scale=0.75,xscale=-1]
\draw[wavy] (-1,0.3) -- (-0.5,0.3) to[out=0,in=0] (-0.5,-0.3) -- (-1,-0.3);
\draw (0,-0.75) -- (0,0.75);
\end{tikzpicture}\label{eq:unitarity}
\end{equation}
and the first coproduct relation (\ref{eq:RcoprJ}) is equivalent to
\begin{equation}\label{pic:RcoprJ}
  \begin{tikzpicture} [baseline=-3pt,scale=0.75]
    \draw (-2,0) -- (2,0);
    \draw (0,-2) node[below]{$R (J_a \otimes 1)$} -- (0,2);
    \draw[wavy=0.3] (-2,1) node[below] {$a$} -- (-1,1) -- (-1,0) node[oper] {};
  \end{tikzpicture}
  +
  \begin{tikzpicture} [baseline=-3pt,scale=0.75]
    \draw (-2,0) -- (2,0);
    \draw (0,-2) node[below]{$R (\thab{a}{b} \otimes J_b)$} -- (0,2);
    \draw[wavy=0.65] (-2,1) node[below] {$a$} -- (-1,1) -- (-1,-1) -- (0,-1) node[oper] {};
  \end{tikzpicture}
  =
  \begin{tikzpicture} [baseline=-3pt,scale=0.75]
    \draw (-2,0) -- (2,0);
    \draw (0,-2) node[below]{$(1 \otimes J_a) R$} -- (0,2);
    \draw[wavy] (-2,1) node[below] {$a$} -- (0,1) node[oper] {};
  \end{tikzpicture}
  +
  \begin{tikzpicture} [baseline=-3pt,scale=0.75]
    \draw (-2,0) -- (2,0);
    \draw (0,-2) node[below]{$(J_b \otimes \thab{a}{b}) R$} -- (0,2);
    \draw[wavy=0.35] (-2,1) node[below] {$a$}
    -- (1,1) -- (1,0) node[oper] {};
  \end{tikzpicture} \quad
\end{equation}
where we recall that the ``time flow''is south-west to north-east.
Similarly, for ``tail operators'' one has
\begin{equation}\label{pic:Rcoprth}
  \begin{tikzpicture} [baseline=-3pt,scale=0.75]
    \draw (-2,0) -- (2,0);
    \draw (0,-2) node[below]{$R (\thab{a}{c} \otimes \thab{c}{b})$} -- (0,2);
    \draw[wavy=0.38] (-1,1) node[left] {$a$} -- (-1,-1) -- (1,-1) node[below] {$b$};
  \end{tikzpicture}
  =
  \begin{tikzpicture} [baseline=-3pt,scale=0.75]
    \draw (-2,0) -- (2,0);
    \draw (0,-2) node[below]{$ (\thab{c}{b} \otimes \thab{a}{c}) R$} -- (0,2);
    \draw[wavy=0.38] (-1,1) node[left] {$a$} -- (1,1) -- (1,-1) node[below] {$b$};
  \end{tikzpicture} \quad
\end{equation}
which means one can move the tail freely across vertices. 
The remaining equations (\ref{eq:Rcoprhth}--\ref{eq:RcoprhJ}) have analogous the graphical equivalents, but the tails now enter from the right:
\begin{equation}\label{pic:Rcoprhth}
\begin{tikzpicture} [baseline=-3pt,scale=0.75]
\draw (-2,0) -- (2,0);
\draw (0,-2) node[below]{$R (\thabhat{b}{c} \otimes \thabhat{c}{a})$}  -- (0,2);
\draw[wavy=0.38] (1,-1) node[below] {$a$} -- (-1,-1) -- (-1,1) node[left] {$b$};
\end{tikzpicture}
=
\begin{tikzpicture} [baseline=-3pt,scale=0.75]
\draw (-2,0) -- (2,0);
\draw (0,-2) node[below]{$(\thabhat{c}{a} \otimes \thabhat{b}{c}) R$}  -- (0,2);
\draw[wavy=0.38] (1,-1) node[below] {$a$} -- (1,1) -- (-1,1) node[left] {$b$};
\end{tikzpicture}
\end{equation}

\begin{equation}\label{pic:RcoprhJ}
  \begin{tikzpicture} [baseline=-3pt,scale=0.75]
    \draw (-2,0) -- (2,0);
    \draw (0,-2) node[below]{$R (1 \otimes \wh J_a)$} -- (0,2);
    \draw[wavy] (2,-1) node[below] {$a$}
    -- (0,-1) node[oper] {};
  \end{tikzpicture}
  +
  \begin{tikzpicture} [baseline=-3pt,scale=0.75]
    \draw (-2,0) -- (2,0);
    \draw (0,-2) node[below]{$R (\wh J_b \otimes \thabhat{b}{a})$} -- (0,2);
    \draw[wavy=0.4] (2,-1) node[below] {$a$}
    -- (-1,-1) -- (-1,0) node[oper] {};
  \end{tikzpicture}
  =
\begin{tikzpicture} [baseline=-3pt,scale=0.75]
    \draw (-2,0) -- (2,0);
    \draw (0,-2) node[below]{$(\wh J_a \otimes 1) R$} -- (0,2);
    \draw[wavy=0.4] (2,-1) node[below] {$a$}
    -- (1,-1) -- (1,0) node[oper] {};
  \end{tikzpicture}
  +
  \begin{tikzpicture} [baseline=-3pt,scale=0.75]
    \draw (-2,0) -- (2,0);
    \draw (0,-2) node[below]{$(\thabhat{b}{a} \otimes \wh J_b) R$} -- (0,2);
    \draw[wavy=0.6] (2,-1) node[below] {$a$}
    -- (1,-1) -- (1,1) -- (0,1) node[oper] {};
  \end{tikzpicture}
\end{equation}

\subsection{Non-local currents and conservation laws in the bulk}
\label{ssec:qg-bulk}
Continuing along the lines of~\cite{BernardFelder91}, we act with the repeated coproduct $\Delta^{(L)}$ on the elements of $U$ to define non-local currents. By iterating the coproduct in \eqref{eq:coprJ}, we obtain
\begin{equation}
\mathbf{J}_a :=
\Delta^{(L)}(J_a)=\sum_{x=1}^L j_a^{(\tsup)}(x),
\quad
j_a^{(\tsup)}(x) :=\delta_{a,a_1}
\thab{a_1}{a_2} \otimes\cdots\otimes \thab{a_{x-1}}{a_x}
\otimes J_{a_x} \otimes 1\otimes\cdots\otimes 1 \, .\label{eq:JLcoprod}
\end{equation}
The object thus constructed, $\mathbf{J}_a$, is the charge associated with the time component $j_a^{(\tsup)}(x)$ of a non-local current. 
Acting on a
tensor product $V_1 \otimes \dots \otimes V_L$, we have the graphical representation
\begin{equation*}
 j^{(\tsup)}_a(x)  = 
  \begin{tikzpicture} [baseline=-3pt,scale=0.75]
    \foreach\x in {1,...,9}
    \draw (\x,1) -- (\x,-1);
    \draw [wavy]
    (0,0) node[below] {$a$} -- (5,0) node[oper] {} ;
    \node at (5,-1.3) {$x$};
  \end{tikzpicture}
\end{equation*}
where each solid line corresponds to a space $V_i$
(the tensor product is ordered from left to right, as in the corresponding
algebraic expression).
In this equation, and in what follows, we use superscripts $\tsup$ and $\xsup$ to indicate
vertical and horizontal directions respectively.  
If $V_i\simeq \mathbb{C}^d$ as below, then each solid line will carry an index
 in $\{1,\ldots,d\}$,
and each wavy line an index $a \in \{1,\ldots,n\}$.
The intersection of a wavy line and a solid line is a $\thab{a_i}{a_{i+1}}$
acting on the solid line. 
It is also useful to introduce the following graphical notation for the sum  occurring
in (\ref{eq:JLcoprod}):
\begin{equation*}
  \mathbf{J}_a=
  \begin{tikzpicture} [baseline=-3pt,scale=0.75]
    \foreach\x in {1,...,9}
    \draw (\x,1) -- (\x,-1);
    \draw[contour=0.75] (0,-0.075) -- (10,-0.075);
    \draw[wavy] (0,0.075) node[left] {$a$} -- (5,0.075) node[oper] {} ;
  \end{tikzpicture}
\end{equation*}
where the dashed line represents summation over the position of the node (\ie, 
discrete integration).


We want to reinterpret ~\eqref{eq:RcoprJ} (or its graphical equivalent, \eqref{pic:RcoprJ}) 
as a discrete current conservation
for a vector field $(j_a^{(\xsup)},j_a^{(\tsup)})$.
This leads us 
to define $j_a^{(\xsup)}(x)$, $x\in\mathbb{Z}+1/2$, 
as the insertion of a dot on the horizontal
edge $[x-1/2,x+1/2]$ (with a tail attached, one half-step up and then to the left). If one wants to 
define $j_a^{(\xsup)}$ as an operator on $V_1\otimes
\cdots\otimes V_L$, one needs to ``embed'' it inside a transfer matrix: \ie, letting $\mathbf{T}$ be the one-row transfer matrix
(recall that
the boundary conditions on the left/right are left undetermined in this
section), 
we define
\begin{equation*}
\mathbf{T}^{1/2}
j^{(\xsup)}_a(x)
\mathbf{T}^{1/2}
=
\begin{tikzpicture}[baseline=-3pt,scale=0.75]
\draw (0,0) -- (8,0);
\foreach\x in {1,...,7}
 \draw (\x,-1) -- (\x,1);
\draw[wavy=0.55]
 (0,0.5) node[below] {$a$} -- (4.5,0.5) -- (4.5,0) node[oper] {} node[below=1mm] {$x$};
\end{tikzpicture}
\end{equation*}
Adding a tail to all terms in ~\eqref{pic:RcoprJ} which extends all the way to the left, we can straighten it using \eqref{pic:Rcoprth}. Assuming that the tail commutes with the left boundary, we find 
\begin{equation} \label{eq:conserv-j}
 j_a^{(\xsup)}(x-1/2,t) - j_a^{(\xsup)}(x+1/2,t)
  + j_a^{(\tsup)}(x,t-1/2) - j_a^{(\tsup)}(x,t+1/2)
  = 0 \,
\end{equation}
where in the operator formalism, the time evolution of any operator $\mathcal{O}$ is given by $\mathcal{O}(t)=\mathbf{T}^t \mathcal{O} \mathbf{T}^{-t}$. Equation \eqref{eq:conserv-j} expresses the conservation of the current $(j_a^{(\xsup)},j_a^{(\tsup)})$ that we have defined.

Summing \eqref{eq:conserv-j} over $x$ results in the conservation law
for the associated global charge $\mathbf{J}_a$ \eqref{eq:JLcoprod}, up to boundary terms:
\begin{equation}\label{eq:conserv-Q}
  \mathbf{J}_a(t+1/2) - \mathbf{J}_a(t-1/2)
  = j_a^{(\xsup)}(1/2, t) - j_a^{(\xsup)}(L+1/2, t) \,.
\end{equation}
If, as above, we depict the charge $\mathbf{J}_a$ by
using a dotted line to denote summation, then equation \eqref{eq:conserv-Q} has the graphical interpretation
\begin{equation*}
  \begin{tikzpicture} [baseline=-3pt,scale=0.75]
    \draw (0,0.25) -- (9,0.25);
    \draw[contour=0.8] (0,0.6) -- (0.35,0.6) -- (0.35,-0.4) -- (8.5,-0.4);
    \foreach\x in {1,...,8}
    \draw (\x,1.25) -- (\x,-0.75);
    \draw [wavy] (0,0.75) node[left] {$a$} -- (0.5,0.75) -- (0.5,-0.25) -- (4,-0.25) node[oper] {} ;
  \end{tikzpicture}
  \quad=\quad
  \begin{tikzpicture} [baseline=-3pt,scale=0.75]
    \draw (0,0.25) -- (9,0.25);
    \draw[contour=0.75] (0,0.6) -- (8.5,0.6) -- (8.5,-0.25) -- (9,-0.25);
    \foreach\x in {1,...,8}
    \draw (\x,1.25) -- (\x,-0.75);
    \draw [wavy] (0,0.75) node[left] {$a$}  -- (5,0.75) node[oper] {} ;
  \end{tikzpicture}
\end{equation*}

In this paper, we shall be mostly concerned with the ``local'' relation
\eqref{eq:conserv-j} and not so much with the global relation
\eqref{eq:conserv-Q}. Note however that even the former relation
is not strictly local because of the tails; we therefore shall have
to pay attention to the boundary conditions to the left when applying
it in what follows. 

It is the purpose of this paper to relate
\eqref{eq:conserv-j}
to the so-called discrete holomorphicity condition. 

\subsection{Adjoint action}
\label{ssec:qg-adjoint}
A detailed discussion of the action of $U$ on local fields 
and of its adjoint action can
be found in \S 2.4 and 2.5 of~\cite{BernardFelder91}. 
Here we summarize some relevant facts. Let us write the general coproduct as
$$
\Delta(X)=\sum X^{(1)} \otimes X^{(2)} \,,
$$
for any $X \in U$.
The adjoint action of a Hopf algebra is defined by
$$
\ad{X}{Y} := \sum X^{(1)} \ Y \ S \left(X^{(2)} \right) \,.
$$
In the case of elements of the form $J_a$, whose coproduct and antipode are given by \eqref{eq:coprJ},
this means that
\begin{equation}
\ad{J_a}{J_b} = J_a J_b - \thab{a}{c} J_b \thabhat{d}{c} J_d \,.\label{eq:adaction}
\end{equation}
The natural action of 
$J_a$ on $\mathbf{J}_b$, viewed as an operator on $V_1\otimes\cdots\otimes V_L$,
is obtained by applying $\Delta^{(L)}$:
\begin{equation*}
  \Delta^{(L)} \left[ \ad{J_a}{J_b} \right] = \sum_{x=1}^{L}
  A_a \left[ j_b^{(\tsup)}(x,t) \right] \,,
\end{equation*}
where $A_a$ on the r.h.s.\ gives the action of $J_a$ on local fields. This action follows from taking the
coproduct of (\ref{eq:adaction})
and is best described graphically:
\begin{eqnarray} \label{eq:adj-graph}
  A_a \left[ j_b^{(\tsup)}(x,t) \right]
  =
  &&\begin{tikzpicture} [baseline=-3pt,scale=0.75]
    \foreach\x in {1,...,8}
    \draw (\x,1) -- (\x,-1);
    \draw[wavy] (0,0) node[left] {$b$} -- (5,0) node[oper] {};
    \node at (5,-1.3) {$x$};
    \draw[contour=0.8] (0,0.5) -- (9.5,0.5);
    \draw[wavy] (0,0.65) node[left] {$a$} -- (6,0.65) node[oper] {};
  \end{tikzpicture}\\
-&&\begin{tikzpicture} [baseline=-3pt,scale=0.75]
    \foreach\x in {1,...,8}
    \draw (\x,1) -- (\x,-1.75);
    \draw[wavy] (0,0) node[left] {$b$} -- (5,0) node[oper] {};
    \node at (5,-1.95) {$x$};
    \draw[contour=0.6]  (0,-1.55) -- (9,-1.55);
    \draw[wavy] (0,0.65) node[left] {$a$} -- (5.65,0.65) --(8.65,0.65) -- 
    (8.65,-0.65) -- (0,-0.65) -- (0,-1.35) -- (3,-1.35) node[oper] {};
  \end{tikzpicture}\nonumber\\
  =
  &&\begin{tikzpicture} [baseline=-3pt,scale=0.75]
    \foreach\x in {1,...,8}
    \draw (\x,1) -- (\x,-1);
    \draw[wavy] (0,0) node[left] {$b$} -- (5,0) node[oper] {};
    \node at (5,-1.3) {$x$};
    \draw[contour=0.8] (0,0.5) -- (5.5,0.5) -- (5.5,-0.5) -- (0,-0.5);
    \draw[wavy] (0,0.65) node[left] {$a$} -- (5.65,0.65) -- 
    (5.65,-0.65) -- (3,-0.65) node[oper] {};
  \end{tikzpicture}\nonumber
\end{eqnarray}
Once again, dashed lines denote discrete contour integration and the final equality follows by application of
relations (\ref{eq:jdual}) and (\ref{eq:unitarity}).

A similar procedure can be carried out for the action on the other component
$j_b^{(\xsup)}$ of currents. In this way, one can organize various currents as multiplets
of $U$ (or of subalgebras of $U$).
In general, starting from some generators $J_a$,
the adjoint action of the whole of $U$
produces a large subspace inside $U$. Note however
that in the case of quantized affine algebras, to be discussed now,
the Serre relations imply that the module generated by the adjoint
action of a $U_q(A_1) := \Uq{sl_2}$ subalgebra on some other
Chevalley generator is finite-dimensional.

We now apply the formalism of previous sections to particular quantized affine algebras.

\subsection{Quantized affine algebras}
\label{ssec:qaa}
In this paper we always take $U$ to be a quantized affine algebra $\Uq{\mathfrak{g}}$ corresponding
to a rank 1 affine Lie algebra $\mathfrak{g}$. Let $\mathcal{A}_{ij}$ denote the entries of the generalized Cartan matrix for
$\Uq{\mathfrak{g}}$, and let $d_i$ be integers such that 
$d_i \mathcal{A}_{ij} = d_j \mathcal{A}_{ji}$, whose greatest common divisor is 1. Then the Chevalley presentation of
$\Uq{\mathfrak{g}}$ is given in terms of the generators $\{E_i,F_i,T_i\}$, $i \in \{0,1\}$,
satisfying the list of relations\footnote{For a review of quantized affine algebras see~\cite{chpr94}.} 
\begin{align}
& T_i T_i^{-1} = T_i^{-1} T_i = 1, \quad [T_i,T_j] = 0 \,,
\label{1}
\\
&
T_i E_j T_i^{-1} = q^{d_i \mathcal{A}_{ij}} E_j, 
\quad 
T_i F_j T_i^{-1} = q^{-d_i \mathcal{A}_{ij}} F_j,
\quad 
[E_i,F_j] = \delta_{ij} \frac{T_i-T_i^{-1}}{q^{d_i}-q^{-d_i}} \,,
\label{2}
\\
&
\sum_{k=0}^{1-\mathcal{A}_{ij}} 
(-)^k \left[ \begin{array}{c} 1-\mathcal{A}_{ij} \\ k \end{array} \right]_{q^{d_i}}
(E_i)^{1-\mathcal{A}_{ij}-k} E_j (E_i)^k = 0 \,,
\label{3}
\\
&
\sum_{k=0}^{1-\mathcal{A}_{ij}} 
(-)^k \left[ \begin{array}{c} 1-\mathcal{A}_{ij} \\ k \end{array} \right]_{q^{d_i}}
(F_i)^{1-\mathcal{A}_{ij}-k} F_j (F_i)^k = 0 \,,
\label{4}
\end{align}
where we use the notation
\begin{align*}
\left[\begin{array}{c} m \\ n \end{array}\right]_q
:=
\frac{(q^m-q^{-m})\dots (q^{m-n+1}-q^{-m+n-1})}{(q^n-q^{-n})\dots (q-q^{-1})} \,.
\end{align*} 
The coproduct of these generators is taken to be 
\begin{align*}
& \Delta(E_i) = E_i \otimes 1 + T_i \otimes E_i \,,& 
&\Delta(F_i) = F_i \otimes T_i^{-1} + 1 \otimes F_i \,,&
\Delta(T_i) = T_i \otimes T_i \,.
\end{align*}
It is also useful for our purposes to introduce the modified generators $\bar E_i := q^{d_i} T_i F_i$, whose coproduct takes the more convenient form: 
\begin{align*}
\Delta(\bar E_i) = \bar E_i \otimes 1 + T_i \otimes \bar E_i \,.
\end{align*}
The correspondence between these generators and the Hopf algebra elements $J_a,\thab{a}{b},\thabhat{a}{b}$ introduced in \secref{ssec:hopf} is immediate. For $a=0,1$ we set: $J_a = E_a, \bar E_a$, their coproduct 
being of the form of \eqref{eq:coprJ}; $\thab{a}{b}= \delta_{a,b} T_a$ (resp.\ $\thabhat{a}{b} = \delta_{a,b} T_a^{-1}$) their coproduct being of the form \eqref{eq:coprth} (resp.\ \eqref{eq:coprhth}); $\wh J_a = F_a$, their coproduct being of the form \eqref{eq:coprhJ}).    

In the following subsections we describe the two quantized affine algebras which will interest us in this paper, as well as the representations to be considered.

\subsubsection{The case $U = U_q(A_1^{(1)})$}
The first case of interest to us is $U_q(A_1^{(1)})$, for which the generalized Cartan matrix is given by
\[ 
\left( 
\begin{array}{cc}
\mathcal{A}_{00} & \mathcal{A}_{01} \\
\mathcal{A}_{10} & \mathcal{A}_{11}
\end{array} 
\right) 
= 
\left( 
\begin{array}{cc}
2 & -2 \\
-2 & 2
\end{array} 
\right) 
\]
and $d_0=d_1=1$.
The representation $(\pi_z,V_z)$ is taken as the level-zero fundamental (principal) evaluation representation $V_z=\mathbb{C}^2[[z]]$
\begin{align*}
  \pi_z(E_0)&= \begin{pmatrix}0&0\\z&0\end{pmatrix}& 
    \pi_z(\bar E_0)&= \begin{pmatrix}0&z^{-1}\\0&0\end{pmatrix}& \pi_z(T_0)&=\begin{pmatrix}q^{-1}&0\\0&q\end{pmatrix}
      \\
      \pi_z(E_1)&=\begin{pmatrix}0&z\\0&0\end{pmatrix}& 
      \pi_z(\bar E_1)&=\begin{pmatrix}0&0\\z^{-1}&0\end{pmatrix}& \pi_z(T_1)&=\begin{pmatrix}q&0\\0&q^{-1}\end{pmatrix} \,.
\end{align*}
The $R$-matrix $R(z)$ is given by 
\begin{equation} \label{eq:R-6V}
  R(z) = \left(\begin{array}{cc|cc}
    qz-q^{-1}z^{-1} & 0 & 0 & 0 \\
    0 & z-z^{-1} & q-q^{-1} & 0 \\
\hline
    0 & q-q^{-1} & z-z^{-1} & 0 \\
    0 & 0 & 0 & qz-q^{-1}z^{-1}
  \end{array}\right) \,,
  \end{equation}
which gives the weights of the 6-vertex model in the principal gradation.

\subsubsection{The case $U=U_q(A_2^{(2)})$}
The second case which we study is $ U_q(A_2^{(2})$, for which the generalized Cartan matrix is
\[ 
\left( 
\begin{array}{cc}
\mathcal{A}_{00} & \mathcal{A}_{01} \\
\mathcal{A}_{10} & \mathcal{A}_{11}
\end{array} 
\right) 
= 
\left( 
\begin{array}{cc}
2 & -4 \\
-1 & 2
\end{array} 
\right) 
\]
and $d_0=1$, $d_1=4$.
The representation $(\pi_z,V_z)$ is now $V_z=\mathbb{C}^3[[z,z^\ell]]$
with 
\begin{align*}
\pi_z(E_0) &= 
z^{1-\ell} \varphi(q)
\left( \begin{array}{ccc} 0 & 0 & 0 \\ 1 & 0 & 0 \\ 0 & q & 0 \end{array} \right)
&
\pi_z(E_1) &= 
z^{+2\ell}
\left( \begin{array}{ccc} 0 & 0 & 1 \\ 0 & 0 & 0 \\ 0 & 0 & 0 \end{array} \right)
\\
\pi_z(\bar E_0) &= 
z^{\ell-1} \varphi(q)
\left( \begin{array}{ccc} 0 & q^{-1} & 0 \\ 0 & 0 & 1 \\ 0 & 0 & 0 \end{array} \right)
&
\pi_z(\bar E_1) &= 
z^{-2\ell}
\left( \begin{array}{ccc} 0 & 0 & 0 \\ 0 & 0 & 0 \\ 1 & 0 & 0 \end{array} \right)
\\
\pi_z(T_0) &= 
\left( \begin{array}{ccc} q^{-2} & 0 & 0 \\ 0 & 1 & 0 \\ 0 & 0 & q^{2} \end{array} \right)
&
\pi_z(T_1) &= 
\left( \begin{array}{ccc} q^{4} & 0 & 0 \\ 0 & 1 & 0 \\ 0 & 0 & q^{-4} \end{array} \right) \,,
\end{align*}
where $\varphi(q) = (q+q^{-1})^{1/2}$, and $\ell$ is an arbitrary constant which we will fix later.
The $R$-matrix is
\begin{equation} \label{eq:R-19V}
  R(z) = \left( \begin{array}{ccc|ccc|ccc}
    \omega_{14} &&&&&&&& \\
    & \omega_{10} && \omega_5 &&&&& \\
    && \omega_{16} && \omega_8 && \omega_{19} \\
    \hline
    & \omega_2 && \omega_{12} &&&&& \\
    && \omega_7 && \omega_1 && \omega_6 \\
    &&&&& \omega_{13} && \omega_3 \\
    \hline
    && \omega_{18} && \omega_9 && \omega_{17} \\
    &&&&& \omega_4 && \omega_{11} \\
    &&&&&&&& \omega_{15}
  \end{array} \right) \,,
\end{equation}
where the entries are given by
\begin{align*}
    \omega_1 &= (z-z^{-1})(q^3z+q^{-3}z^{-1}) + (q^2-q^{-2})(q^3+q^{-3}) \\
    \omega_2 = \omega_4 &= z^{-\ell} (q^2-q^{-2})(q^3z+q^{-3}z^{-1}) \\
    \omega_3 = \omega_5 &= z^{+\ell} (q^2-q^{-2})(q^3z+q^{-3}z^{-1}) \\
    \omega_6 = \omega_8 &= -q^{+2} z^{+\ell} (q^2-q^{-2})(z-z^{-1}) \\
    \omega_7 = \omega_9 &= +q^{-2} z^{-\ell} (q^2-q^{-2})(z-z^{-1}) \\
    \omega_{10} = \omega_{11} = \omega_{12} = \omega_{13}
    &= (z-z^{-1})(q^3 z + q^{-3} z^{-1}) \\
    \omega_{14} = \omega_{15} &= (q^2z-q^{-2}z^{-1})(q^3z+q^{-3}z^{-1}) \\
    \omega_{16} = \omega_{17} &= (z-z^{-1})(qz+q^{-1}z^{-1}) \\
    \omega_{18} &= z^{-2\ell} (q^2-q^{-2}) [(q^2+q^{-2})qz^2 - (q-q^{-1}) q^{-2}] \\
    \omega_{19} &= z^{+2\ell} (q^2-q^{-2}) [(q^2+q^{-2}) q^{-1} z^{-2} + (q-q^{-1})q^2] \,.
\end{align*}
These coincide with the Boltzmann weights of the Izergin--Korepin 19-vertex model, up to factors of $z^{\pm\ell}$ which come from the fact that we have not yet fixed the gradation. 




\section{From vertex models to loop models}
\label{sec:vertex-loop}
In this section we review the loop/vertex  model connection for  $U_q(A_1^{(1)})$ and $U_q(A_2^{(2)})$ models in the bulk. Until this stage, our analysis has been on a lattice $\Omega$ with an arbitrary angle between horizontal and vertical lines. We now specify our domain further, by considering a rhombic lattice of definite angle $\alpha$, which we will ultimately relate to the ratio of the spectral parameters $z/w$. We also draw the dual lattice using dotted lines:
\begin{equation}\label{eq:plaq}
R=
\begin{tikzpicture}[baseline=-3pt,scale=0.75,distort]
\draw (1,0) -- (-1,0) node[left] {$z$};
\draw (0,1) -- (0,-1) node[below] {$w$};
\draw (-0.3,0) arc (180:90:0.3) node[left=1.5mm] {$\alpha$};
\end{tikzpicture}
=
 \begin{tikzpicture}[baseline=-3pt,scale=1.25,distort]
\plaq (0,0)
 \end{tikzpicture}
\end{equation}
where the edges of an elementary ``plaquette'' (elementary rhombus
of the dual lattice) are of unit length.

\subsection{The dense \TL{} model and the \texorpdfstring{$U_q(A_1^{(1)})$}{Uq(A1(1))} vertex model}
The dense {\TL} model is defined by assigning weights $a$ and $b$ to the following two local configurations of a rhombus with top-left lattice angle $\alpha$:
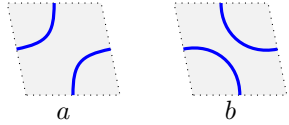
\begin{center}
 \begin{tabular}{ccc}
  \begin{tikzpicture} [baseline=-3pt,scale=1.25,distort]
    \plaqa(0,0)
  \end{tikzpicture}
  & &
  \begin{tikzpicture} [baseline=-3pt,scale=1.25,distort]
    \plaqb(0,0)
  \end{tikzpicture}
\\
$a$ & & $b$ 
  \end{tabular}
\end{center}
and weight $\tau$ to closed loops in a given lattice  configuration $C$.
Thus, $C$  is assigned the Boltzmann weight
\begin{equation} \label{eq:W-TL}
  W(C) = a^{N_a(C)} \ b^{N_b(C)} \ \tau^{N_\ell(C)} \,,
\end{equation}
where $N_a(C)$ (resp.\ $N_b(C)$) is the number of plaquettes of weight $a$ (resp.\ $b$),
and $N_\ell(C)$ is the number of closed loops in $C$.

Let us now recall how~\cite{Bax82} these weights are related to those of the six-vertex model.
Firstly, we identify
\begin{equation} 
\tau=-(q+q^{-1}),
\quad\quad 
q:= -e^{2\pi i\nu} .
\end{equation} 
We then associate the above configurations with operators $A$, $B: V\ot V \ra V\ot V$ (acting SW-NE),
with $V=\mathbb{C} \ket{\uparrow}\oplus \mathbb{C} \ket{\downarrow}$. This is done by dressing each configuration with arrows and reading off the associated 
weight $e^{i\nu \delta}$ associated with total turning angle $\delta$.\footnote{
  We assume, as shown on the pictures, 
  that the loop lines
  enter/leave {\em orthogonally}\/ to the sides of the rhombus.} In this way we obtain
  \begin{equation}\label{eq:TL-gen}
  A := \left( \begin{array}{cc|cc}
    1 & 0 & 0 & 0 \\
    0 & e^{2i\nu\alpha} & 0 & 0 \\
    \hline
    0 & 0 & e^{-2i\nu\alpha} & 0 \\
    0 & 0 & 0 & 1
  \end{array} \right) \,,
  \qquad
  B:= \left( \begin{array}{cc|cc}
    0 & 0 & 0 & 0 \\
    0 & e^{-2i\nu(\pi-\alpha)} & 1 & 0 \\
    \hline
    0 & 1 & e^{2i\nu(\pi-\alpha)} & 0 \\
    0 & 0 & 0 & 0
  \end{array} \right) \,,
\end{equation} 
in the basis $\{ \ket{\uparrow\uparrow}, \ket{\uparrow\downarrow},
\ket{\downarrow\uparrow}, \ket{\downarrow\downarrow} \}$. 
The $\cR$-matrix is then the linear combination of these operators, dressed by their respective weights:
\begin{equation} \label{eq:R-6V-loop}
  \Rc := P R = a\ A + b\ B
  = \left( \begin{array}{cc|cc}
    a & 0 & 0 & 0 \\
    0 & e^{2i\nu\alpha}a + e^{-2i\nu(\pi-\alpha)}b & b & 0 \\
\hline
    0 & b & e^{-2i\nu\alpha}a + e^{2i\nu(\pi-\alpha)}b & 0 \\
    0 & 0 & 0 & a
  \end{array} \right) \,.
\end{equation}

\subsection{Six-vertex model weights from intertwining relations}
It is well known that the commutant of the quantized algebra
$\Uq{A_1}$ acting on a tensor product of two-dimensional representations
is the {\TL} algebra~\cite{Jim85}.
This suggests that the two terms $A$ and $B$
in~\eqref{eq:R-6V-loop} are intertwiners for the subalgebra
$\aver{E_1,\bar{E}_1,T_1} \cong \Uq{A_1}$. Indeed, one can check that
\begin{equation}\label{eq:tl-AB}
  \begin{cases}
    A\, \pi_{z,w} (\Delta(X)) = \pi_{w,z} (\Delta(X)) \, A \\
    B\, \pi_{z,w}(\Delta(X)) = \pi_{w,z} (\Delta(X)) \, B  \,,
  \end{cases}
\qquad
   X =E_1,\bar{E}_1,T_1 \,,
  \qquad
\end{equation}
provided the following relation holds between angle and spectral parameters:
$z/w=e^{-2i\nu\alpha}$.
Notice that the equation for $X=T_1$ is automatically satisfied,
since both $A$ and $B$ preserve the total magnetization.

Moreover, imposing that $\check R$ commute in a similar way with the action of $E_0$, $\bar E_0$
fixes the ratio $a/b$, say 
\begin{equation}
a=q\,x-q^{-1}x^{-1},\quad\quad b=x-x^{-1}, \quad\quad x:= z/w.
\end{equation}
Substituting these values for the weights into \eqref{eq:R-6V-loop}, we recover the 6-vertex $R$-matrix \eqref{eq:R-6V}. 

{\em Remark:} The explicit connection of the operators \eqref{eq:TL-gen} with the usual generators of the Temperley--Lieb algebra is as follows. On a strip of width $L$, the {\TL} algebra with generators $\{g_1, \dots, g_L\}$ satisfies the list of relations
\begin{equation}
  \begin{array}{rcl}
    g_j g_{j \pm 1} g_j &=& g_j \\
    g_j^2 &=& \tau g_j \\
    g_j g_k &=& g_k g_j, \quad \text{if $|j-k|>1$.}
  \end{array}
\end{equation}
The generator $g_j$ acts non-trivially on spaces at positions $j$ and $j+1$, and the weights for a plaquette at this position are encoded in the $\check{\mathcal R}$-matrix $\check{\mathcal R}_{j,j+1} = a\ \myid + b\ g_j$. Both the identity $\myid$ and TL generator $g_j$ have well known graphical interpretations, as a 45 degree rotation and deformation of the tiles above into squares.

Consequently, we expect that $\cR$ and $\check{\mathcal R}$ are related by a simple gauge transformation (corresponding to the passage from principal
gradation to homogeneous gradation), which is indeed the case; we find that
$$
\check{\mathcal R} = \mathcal{U}^{-1} \ \Rc \ \mathcal{U}' \,,
\qquad
\mathcal{U} := w^{\sigma^z/2} \otimes z^{\sigma^z/2} \,,
\qquad
\mathcal{U}' := z^{\sigma^z/2} \otimes w^{\sigma^z/2} \,,
$$
since under this transformation the two terms in $\cR$ become
$$
\mathcal{U}^{-1} \ A \ \mathcal{U}' = \myid \,,
\qquad
\mathcal{U}^{-1} \ B \ \mathcal{U}' = \left( \begin{array}{cccc}
    0 & 0 & 0 & 0 \\
    0 & -q^{-1} & 1 & 0 \\
    0 & 1 & -q & 0 \\
    0 & 0 & 0 & 0
  \end{array} \right) \,,
$$
where the second term is the well-known spin-$\half$ representation of $g_j$.

\subsection{The dilute \TL{} model and the 
\texorpdfstring{$U_q(A_2^{(2)})$}{Uq(A2(2))}
vertex model}
The dilute {\TL} model (or $\On$ model) is defined
by the plaquette configurations:
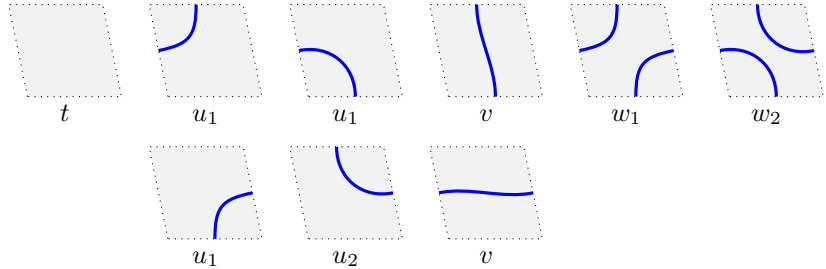
\begin{center}
  \begin{tabular}{cccccccc}
    \begin{tikzpicture} [baseline=-3pt,scale=1.25,distort]
      \plaqg(0,0)
    \end{tikzpicture}
    &
    \begin{tikzpicture} [baseline=-3pt,scale=1.25,distort]
      \plaqf(0,0)
    \end{tikzpicture}
    &
    \begin{tikzpicture} [baseline=-3pt,scale=1.25,distort]
      \plaqd(0,0)
    \end{tikzpicture}
    &
    \begin{tikzpicture} [baseline=-3pt,scale=1.25,distort]
      \plaqc(0,0)
    \end{tikzpicture}
    &
    \begin{tikzpicture} [baseline=-3pt,scale=1.25,distort]
      \plaqa(0,0)
    \end{tikzpicture}
    &
    \begin{tikzpicture} [baseline=-3pt,scale=1.25,distort]
      \plaqb(0,0)
    \end{tikzpicture}\\
%
    $t$ & $u_1$&$u_1$ &$v$ & $w_1$ & $w_2$ \\[2mm] 
    &
    \begin{tikzpicture} [baseline=-3pt,scale=1.25,distort]
      \plaqff(0,0)
    \end{tikzpicture}
    &
    \begin{tikzpicture} [baseline=-3pt,scale=1.25,distort]
      \plaqdd(0,0)
    \end{tikzpicture}
    &
    \begin{tikzpicture} [baseline=-3pt,scale=1.25,distort]
      \plaqcc(0,0)
    \end{tikzpicture} \\
    & $u_1$ & $u_2$ & $v$
  \end{tabular}
\end{center}
with corresponding weights shown beneath the configurations. 
The Boltzmann weight of a configuration $C$ is given by
$$
W(C) = t^{N_t(C)} \ u_1^{N_{u_1}(C)}  u_2^{N_{u_2}(C)} w_1^{N_{w_1}(C)} w_2^{N_{w_2}(C)}
\tau^{N_\ell(C)} \,,
$$
where $N_\alpha$ is the number of plaquettes of type $\alpha$, and $N_\ell$ is the number
of closed loops.
In a similar way to the dense case, we introduce the parameters
$$\tau = -(q^4+q^{-4}), \quad\quad
q := e^{i\pi(\frac{\nu}{2}-\frac{1}{4})} .
$$
Now we identify the configurations with operators $T,U'_1,U''_1,U'_2,U''_2,V',V'',W_1,W_2 :$ $V\ot V \ra V\ot V$, where $V=\mathbb{C}^3=\mathbb{C}\ket{\uparrow}\oplus \mathbb{C} \ket{0} \oplus \mathbb{C}\ket{\downarrow}$. We dress lines in the plaquette with arrows and associate them with $\ket{\uparrow}$ or $\ket{\downarrow}$, we identify missing lines with $\ket{0}$, and collect an associated weight $e^{i\nu\delta}$ for the total turning angle $\delta$. To save space, we will not write down the explicit form of the resulting operators. 

The $\cR$-matrix is, as before, the linear combination of all operators dressed by their Boltzmann weights. In the basis 
$\{\ket{\uparrow\uparrow},\ket{\uparrow 0},\ket{\uparrow\downarrow},
\ket{0\uparrow},\ket{00},\ket{0\downarrow},
\ket{\downarrow\uparrow},\ket{\downarrow 0},\ket{\downarrow\downarrow}\}$, it is given by
\begin{multline}
\label{eq:R-19V-loop}
\cR
=
t\ T + u_1\ (U'_1 + U''_1) + u_2\ (U'_2 + U''_2) + v\ (V' + V'') + w_1\ W_1 + w_2\ W_2
=
\\
\left(
\begin{array}{ccc|ccc|ccc}
w_1 & 0 & 0 & 0 & 0 & 0 & 0 & 0 & 0
\\
0 & u_1 e^{i\nu\alpha} & 0 & v & 0 & 0 & 0 & 0 & 0
\\
0 & 0 & (w_1 + w_2 e^{-2i\nu\pi}) e^{2i\nu\alpha} & 0 & u_2 e^{-i\nu(\pi-\alpha)} & 0 & w_2 & 0 & 0
\\
\hline
0 & v & 0 & u_1 e^{-i\nu\alpha} & 0 & 0 & 0 & 0 & 0
\\
0 & 0 & u_2 e^{-i\nu (\pi-\alpha)} & 0 & t & 0 & u_2 e^{i\nu (\pi-\alpha)} & 0 & 0
\\
0 & 0 & 0 & 0 & 0 & u_1 e^{i\nu\alpha} & 0 & v & 0
\\
\hline
0 & 0 & w_2 & 0 & u_2 e^{i\nu (\pi-\alpha)} & 0 & (w_1 + w_2 e^{2i\nu\pi})e^{-2i\nu\alpha}  & 0 & 0
\\
0 & 0 & 0 & 0 & 0 & v & 0 & u_1 e^{-i\nu\alpha} & 0
\\ 
0 & 0 & 0 & 0 & 0 & 0 & 0 & 0 & w_1
\end{array}
\right)
\end{multline}
   
\subsection{Nineteen-vertex model weights from intertwining relations}
In analogy with the dense TL case, we wish to identify the plaquette operators as intertwiners of the subalgebra $\langle E_1, \bar E_1, T_1 \rangle$. We find that
$$
Y \, \pi_{z,w}(\Delta(X)) = \pi_{w,z}( \Delta(X)) \, Y \,,
\qquad
 X = E_1,\bar E_1, T_1 \,, 
\quad Y= T,U'_1,U''_1,U'_2,U''_2,V',V'',W_1,W_2 \,,
$$
provided we set $(z/w)^{2\ell} = e^{-2i\nu\alpha}$. Furthermore, imposing that
$$
\cR \ \pi_{z,w} (\Delta(X)) = \pi_{w,z} (\Delta(X)) \ \cR
\qquad X= E_0,\bar E_0, T_0 \,,
$$
determines the Boltzmann weights up to normalization:
\begin{align*} 
    t &= (x - x^{-1})(q^3 x + q^{-3} x^{-1}) + (q^2-q^{-2})(q^3+q^{-3}) \\
    u_1 &= (q^{2}-q^{-2}) (q^3 x + q^{-3} x^{-1}) \\
    u_2 &= i (q^{2}-q^{-2}) (x - x^{-1}) \\
    v &= (x - x^{-1})(q^3 x + q^{-3} x^{-1}) \\
    w_1 &= (q^2 x - q^{-2} x^{-1})(q^3 x + q^{-3} x^{-1}) \\
    w_2 &= (x - x^{-1})(q x + q^{-1} x^{-1}) \,,
\end{align*}
where we have again set $x:=z/w$. Inserting these values for the weights into~\eqref{eq:R-19V-loop},
we recover the 19-vertex $R$-matrix~\eqref{eq:R-19V}. 

\section{Discrete holomorphicity in the bulk}
\label{sec:DH-bulk}
The goal of the present section is to connect
\eqref{eq:conserv-j}
to discrete holomorphicity. To do so, we must finally specify our boundary conditions. In all cases to be considered, we shall choose ``reflecting'' boundary conditions, in which adjacent edges on the boundary are paired by a common loop (or are empty, which is allowed in the dilute TL model). The only exception to this rule will be two (or one, in the dilute case) boundary edges from which an open, unpaired loop propagates. 

The reason for making this choice is that these are simple boundary conditions for which the boundary trivially commutes with the tail operators $T_0,T_1$, allowing us to apply \eqref{eq:conserv-j}. Indeed, our results extend to any choice of boundaries for which the tail operators satisfy this property. In particular, we wish to emphasise that our results \emph{do not require integrable boundary conditions.} 

The discussion of integrable boundaries, and the associated boundary discrete holomorphicity, is deferred to \S \ref{sec:vertex-bound}--\ref{sec:DH-bound}.

\subsection{Application to the dense loop model}\label{ssec:loop-dense}
\subsubsection{Loop observables associated to $E_0$ and $\bar E_0$}
Let us now consider the insertion on the lattice of the current $e_0$ 
associated to the operator $E_0$.
We want to translate this insertion into the language of loops.
We consider a model in which all loops are closed except one open path $\gamma$ that
connects two fixed boundary defects. In the vertex model,
this is simply done by setting free boundary conditions for the two boundary defects. An example of such a configuration is shown below.

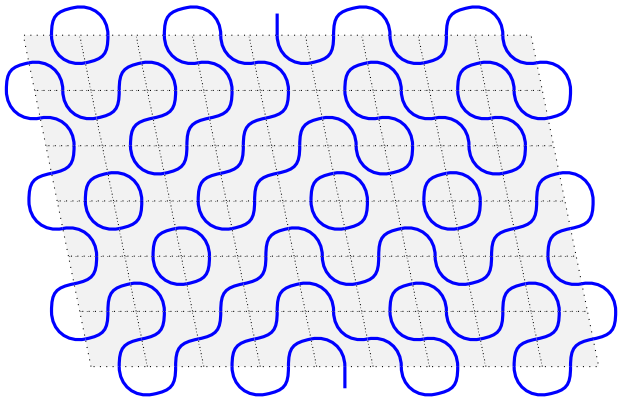
\begin{center}
\begin{tikzpicture}[scale=0.75,distort]
\foreach\x in {0,...,8}
\foreach\y in {0,...,5}
{\pgfmathrandominteger{\rand}{0}{1}
\ifnum\rand=0\plaqa(\x,\y)\else\plaqb(\x,\y)\fi
}
\foreach\y in {0,2,4}
{
        \draw[edge] (-0.5,\y) \start\west\go\north\go\east;
        \draw[edge] (8.5,\y) \start\east\go\north\go\west;
}
\foreach\x in {0,2,5,7}
{
        \draw[edge] (\x,5.5) \start\north\go\east\go\south;
        \draw[edge] (\x,-0.5) \start\south\go\east\go\north;
}
\draw[edge] (4,-0.5) -- ++(\as:0.4);
\draw[edge] (4,5.5) -- ++(\an:0.4);
\end{tikzpicture}
\end{center}


The important observation is that if the marked edge on which $E_0$ sits
belongs to a closed loop, then the contribution is necessarily zero. This is because
$E_0$ changes the direction of an arrow on a dressed loop configuration, whereas
the direction of arrows is continuous around a closed loop.
Graphically,
\begin{equation*}
  \begin{tikzpicture} [baseline=-3pt,scale=0.6]
    \draw[edge] (0,0) arc (0:360:0.8cm);
    \node at (0.6,0) {$E_0$};  
    \draw[wavy] (-2.4,0) -- (0,0) node[oper] {};
  \end{tikzpicture}
  \ =0,
  \qquad \text{and} \qquad
  \begin{tikzpicture} [baseline=-3pt,scale=0.6]
    \draw[edge] (0,0) arc (0:360:0.8cm); 
    \node at (-1,0) {$E_0$};
    \draw[wavy] (-4,0) -- (-1.6,0) node[oper] {};
  \end{tikzpicture}
  = 0 \,.
\end{equation*}
To be precise, we have included the tail in these equations, but in fact the tail has an irrelevant contribution, since it is diagonal in the evaluation representation we are using. In subsequent pictures, the node marking will always correspond to insertion of
$E_0$, and the wavy line to $T_0$.
We conclude that in order to have a non-zero contribution,
the marked edge must lie on the open path $\gamma$, and that the latter, whose
orientation we had left unspecified so far, must be incoming at both
boundaries:

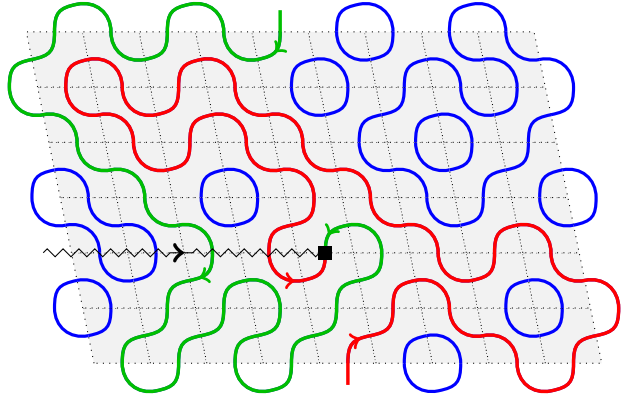
\begin{center}
\slow{
\begin{tikzpicture} [scale=0.75,distort]
\pgfmathsetseed{21}
\foreach\x in {0,...,8}
\foreach\y in {0,...,5}
{\pgfmathrandominteger{\rand}{0}{1}
\ifnum\rand=0\plaqb(\x,\y)\else\plaqa(\x,\y)\fi
}
\foreach\y in {0,2,4}
{
  \draw[edge] (-0.5,\y) \start\west\go\north\go\east;
  \draw[edge] (8.5,\y) \start\east\go\north\go\west;
}
\foreach\x in {0,2,5,7}
{
  \draw[edge] (\x,5.5) \start\north\go\east\go\south;
  \draw[edge] (\x,-0.5) \start\south\go\east\go\north;
}
\draw[red,line width=1.3pt,
decoration={markings, mark = at position 0.035 with {\arrow{>}}},
postaction={decorate}] 
($(4,-0.5)-(\an:0.4)$) -- (4,-0.5) \start\north \go\east\go\north\go\east\go\south\go\east\go\south\go\east\go\north \go\east\go\north\go\west \go\north\go\west\go\south\go\west\go\north\go\west\go\north\go\west\go\north\go\west\go\north\go\west\go\south\go\west\go\north\go\west\go\south\go\east\go\south\go\east coordinate(temp);
\draw[red,line width=1.3pt,
decoration={markings, mark = at position 0.88 with {\arrow{>}}},
postaction={decorate}]
(temp) \start\east
\go\north\go\east\go\south\go\east\go\south\go\west\go\south\go\east\go\north;
\draw[green!75!black,line width=1.3pt,
decoration={markings, mark = at position 0.06 with {\arrow{>}}},
postaction={decorate}] 
($(4,5.5)-(\as:0.4)$) -- (4,5.5) \start\south \go\west\go\north\go\west\go\south \go\west\go\north\go\west\go\south \go\west \go\south\go\east \go\south\go\east\go\south\go\east coordinate(temp);
\draw[green!75!black,line width=1.3pt,
decoration={markings, 
mark = at position 0.09 with {\arrow{>}},
mark = at position 0.98 with {\arrow{>}}},
postaction={decorate}] (temp) \start\east
\go\south\go\west\go\south\go\west\go\south \go\east\go\north \go\east\go\north\go\east\go\south\go\west\go\south \go\east\go\north \go\east\go\north\go\east\go\north\go\west\go\south;
\draw[wavy] (-1,1.5) -- (4,1.5) node[oper] {};
\end{tikzpicture}
}
\end{center}

We first discuss $e_0^{(\tsup)}$, \ie, insertion on a horizontal
edge of a plaquette, as on the picture above. We have $\pi_w(E_0)=w\sigma^-$,
and the Boltzmann weight $W(C)$ is the same as in~\eqref{eq:W-TL}.
The open path $\gamma$ however has an additional factor 
$e^{2i\nu\theta(C)}$ where $\theta(C)$ is the angle spanned by 
the portion of $\gamma$ between a boundary entry point and the insertion of $E_0$
(the red or green directed line) -- the two angles are equal. In fact it is easy to see that
$\theta(C)=\pi k(C)$ where $k(C)$ is an integer whose parity is fixed by the
relative locations of marked edge and boundary entry points of the arc
(on the example $k(C)=2$).
Finally, there is a contribution from the tail of $q^{-\sigma^z}$ (wavy line).
It is easy to check that if $k$ is even, red and green lines each cross
(algebraically) the wavy line $k/2$ times,
whereas if $k$ is odd the red line crosses it $\lfloor k/2\rfloor$ times
and the green line $\lceil k/2 \rceil$ times.
In both cases we find that this produces a factor $q^{k(C)}$.
Using $q=e^{i\pi (2\nu-1)}$
and putting everything together we find
\begin{equation*}
  \aver{e_0^{(\tsup)}(x, t)}
  = \frac{w}{Z} \sum_{C | (x, t)\in \gamma} W(C) \ e^{i(4\nu-1)\theta(C)} \,,
\end{equation*}
where we use the notation $\aver{\mathcal{O}}$ for the expectation value of an operator $\mathcal{O}$, and $Z$ is the partition function, $Z=\sum_{C} W(C)$.

Now let us apply the same reasoning to $e_0^{(\xsup)}$, for which a typical configuration is shown below:
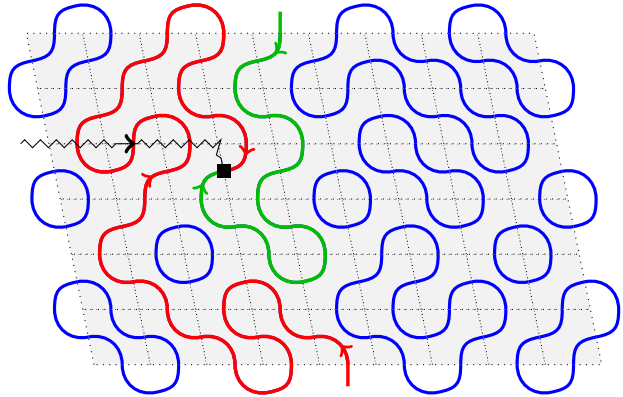
\begin{center}
\slow{
\begin{tikzpicture} [scale=0.75,distort]
\pgfmathsetseed{2}
\foreach\x in {0,...,8}
\foreach\y in {0,...,5}
{\pgfmathrandominteger{\rand}{0}{1}
\ifnum\rand=0\plaqb(\x,\y)\else\plaqa(\x,\y)\fi
}
\foreach\y in {0,2,4}
{
  \draw[edge] (-0.5,\y) \start\west\go\north\go\east;
  \draw[edge] (8.5,\y) \start\east\go\north\go\west;
}
\foreach\x in {0,2,5,7}
{
  \draw[edge] (\x,5.5) \start\north\go\east\go\south;
  \draw[edge] (\x,-0.5) \start\south\go\east\go\north;
}
\draw[red,line width=1.3pt,
decoration={markings, mark = at position 0.03 with {\arrow{>}},
 mark = at position 0.5 with {\arrow{>}},
 mark = at position 0.98 with {\arrow{>}}},
 postaction={decorate}] ($(4,-0.5)-(\an:0.4)$) -- (4,-0.5) \start\north \go\west\go\north\go\west\go\south\go\east\go\south
 \go\west\go\north \go\west\go\north\go\west\go\north\go\east\go\north\go\east\go\north\go\west\go\south\go\west\go\north\go\east\go\north\go\east\go\north
\go\east\go\south \go\west\go\south\go\east\go\south\go\west;
\draw[green!75!black,line width=1.3pt,
decoration={markings, mark = at position 0.07 with {\arrow{>}},
 mark = at position 0.96 with {\arrow{>}}},
 postaction={decorate}] ($(4,5.5)-(\as:0.4)$) -- (4,5.5) \start\south\go\west\go\south\go\east\go\south\go\west\go\south\go\east\go\south\go\west\go\north\go\west\go\north\go\east;
\draw[wavy] (-1,3.5) -- (2.5,3.5) -- (2.5,3) node[oper] {};
\end{tikzpicture}
}
\end{center}
First we have a factor $z$. Then we have a factor from
the angle $e^{2i\nu\theta}$ where this time $\theta=-\alpha+k\pi$ where
$k$ has fixed parity (on the example, $k=-1$). Finally,
the tail crosses algebraically the red line
$\lfloor k/2\rfloor$ times, and the green line $\lceil k/2\rceil$ times.
In total we get $e^{2i\nu\theta}q^{(\theta+\alpha)/\pi}=e^{i(4\nu-1)\theta}e^{i(2\nu-1)\alpha}$,\footnote{If we instead defined $q = e^{i\pi (2\nu+1)}$ we would obtain the total factor $e^{i(4\nu-1)\theta}e^{i(2\nu+1)\alpha}$, leading to an observable which is discretely anti-holomorphic. There is no preferred definition for $q$, since exactly half of the observables are anti-holomorphic regardless of which choice we make.} or
\begin{equation*}
  \aver{e_0^{(\xsup)}(x,t)} = \frac{z \ e^{i(2\nu-1)\alpha}}{Z}
  \sum_{C| (x,t) \in \gamma} W(C) \ e^{i(4\nu-1)\theta(C)} \,.
\end{equation*}
Note that $z \ e^{2i\nu\alpha}=w$.
Therefore, if we define a function on the edges of the lattice
\begin{equation*}
  \phi_0(x,t) := w^{-1}
  \begin{cases}
    e_0^{(\tsup)}(x,t), & (x,t)\in\mathbb{Z}\times(\mathbb{Z}+\half) \\ 
    e^{i\alpha} e_0^{(\xsup)}(x,t), & (x,t)\in(\mathbb{Z}+\half)\times\mathbb{Z} \,, 
  \end{cases}
\end{equation*}
then we have
\begin{equation*}
  \aver{\phi_0(x,t)} = \frac{1}{Z} \sum_{C| (x,t) \in \gamma} W(C) \ e^{i(4\nu-1)\theta(C)} \,,
\end{equation*}
where $\theta(C)$ is the angle spanned by the open path $\gamma$ from one boundary
to $(x,t)$.

Furthermore, the current conservation equation \eqref{eq:conserv-j} for $j_a = e_0$ is 
\begin{equation}
\label{eq:conserv-e}
  e_0^{(\xsup)}(x-1/2,t)
  - e_0^{(\xsup)}(x+1/2,t)
  + e_0^{(\tsup)}(x,t-1/2)
  - e_0^{(\tsup)}(x,t+1/2) = 0 \,.
\end{equation}
Rewriting \eqref{eq:conserv-e} in terms of $\phi_0(x,t)$ we find
\begin{equation} \label{eq:phi0_dh}
  \phi_0(x,t-1/2)
  +e^{i(\pi-\alpha)}\phi_0(x+1/2,t)
  -e^{i(\pi-\alpha)}\phi_0(x-1/2,t)
  -\phi_0(x,t+1/2) = 0 \,,
\end{equation}
which is precisely the discrete holomorphicity condition
around a plaquette of the type:
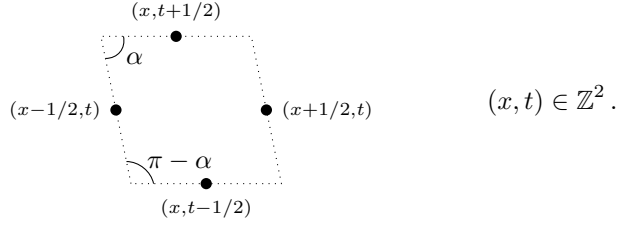
\begin{equation*}
  \begin{tikzpicture} [distort,baseline=1cm]
    \draw[dotted] (0,0) rectangle (2,2);
    \draw (0.3,0) arc (0:90:0.3) node[right=1.5mm] {$\pi-\alpha$};
    \draw (0.3,2) arc (0:-90:0.3) node[right=1.5mm] {$\alpha$};
    \node[circle,fill,inner sep=1.5pt,label={below:$\ss(x,t-1/2)$}] at (1,0) {};
    \node[circle,fill,inner sep=1.5pt,label={right:$\ss(x+1/2,t)$}] at (2,1) {};
    \node[circle,fill,inner sep=1.5pt,label={above:$\ss(x,t+1/2)$}] at (1,2) {};
    \node[circle,fill,inner sep=1.5pt,label={left:$\ss(x-1/2,t)$}] at (0,1) {};
  \end{tikzpicture}
  \qquad\qquad (x,t)\in\mathbb{Z}^2 \,.
\end{equation*}
This equality is valid at the operator level, \ie, when inserted in an arbitrary correlation function
$\aver{\cdots}$.

Note that $\phi_0$ is exactly the lattice holomorphic observable
identified in~\cite{RivaC06,IkhlefC09}: we have shown here that this
observable can be obtained by the construction of conserved currents
associated to the $\Uq{A_1^{(1)}}$ symmetry of~\cite{BernardFelder91}.

Everything can be repeated with $\bar E_0$ instead of $E_0$. This leads to
the conjugate observable
\begin{equation*}
  \aver{\bar\phi_0(x,t)} = \frac{1}{Z} \sum_{C| (x,t) \in \gamma} W(C) \ e^{-i(4\nu-1)\theta(C)} \,,
\end{equation*}
which therefore satisfies an
antiholomorphicity condition.


\subsubsection{Loop observables associated to $E_1$ and $\bar E_1$}
There is a simpler observable, obtained by considering $E_1$.
We use the same type of boundary conditions as above; the whole discussion
goes through, except for the following modifications.
Compared to $E_0$, the arrows on the open path $\gamma$
are inverted, but the tail is inverted as well (it is made
of $q^{+\sigma^z}$). The result
is that the tail and the angle factor almost compensate; for $e_1^{(\tsup)}$
we find using the exact same reasoning that
\begin{equation*}
  \aver{e_1^{(\tsup)}(x,t)} = \frac{w}{Z} \sum_{C| (x,t) \in \gamma} W(C) \ e^{-i\theta(C)} \,.
\end{equation*}
Note that $e^{-i\theta}=(-1)^k$ is independent of the configuration
and simply measures the flux of $\gamma$ through the marked edge 
(\ie, the only configurations that contribute to $\langle e_1^{(\tsup)} \rangle$
are those for which the average flux is nonzero, that is,
where the open path goes through the marked edge).
Similarly, for $e_1^{(\xsup)}$ one finds $e^{-2i\nu\theta} q^{(\theta+\alpha)/\pi}=
e^{-i\theta}e^{i(2\nu-1)\alpha}
$, so
\begin{equation*}
  \aver{e_1^{(\xsup)}(x,t)} = \frac{z \ e^{i(2\nu-1)\alpha}}{Z}
  \sum_{C| (x,t) \in \gamma} W(C) \ e^{-i\theta(C)} \,.
\end{equation*}
Again the same miracle happens in that $z e^{2i\nu\alpha}=w$.
Note that the conservation law for $e_1$ is the obvious conservation of the flux
of $\gamma$ through a plaquette.

Finally, we define
\begin{equation*}
  \phi_1(x,t) := w^{-1}
  \begin{cases}
    e_1^{(\tsup)}(x,t), & (x,t)\in\mathbb{Z}\times(\mathbb{Z}+\frac{1}{2}) \\
    e^{i\alpha} e_1^{(\xsup)}(x,t), & (x,t)\in(\mathbb{Z}+\half)\times\mathbb{Z} \,,
  \end{cases}
\end{equation*}
and we have
\begin{equation*}
 \aver{\phi_1(x,t)} = \frac{1}{Z} \sum_{C| (x,t) \in \gamma} W(C) \ e^{-i\theta(C)} \,,
\end{equation*}
with the discrete holomorphicity equation
\begin{equation} \label{eq:phi1_dh}
  \phi_1(x,t-1/2)
  +e^{i(\pi-\alpha)}\phi_1(x+1/2,t)
  -e^{i(\pi-\alpha)}\phi_1(x-1/2,t)
  -\phi_1(x,t+1/2) = 0\,.
\end{equation}
Note that since $E_1$ commutes with each loop plaquette separately,
\cf \eqref{eq:tl-AB}, the equation above is actually valid for each individual
configuration of the loop model.

Similarly, insertion of $\bar E_1$ leads to the antiholomorphic observable
$$
\aver{\bar\phi_1(x,t)} = \frac{1}{Z} \sum_{C| (x,t) \in \gamma} W(C) \ e^{+i\theta(C)} \,.
$$

In all cases of observables, we observe
that the different Borel subalgebras ($E$ operators versus $\bar E$ operators)
correspond to different chiralities from the point of view of Conformal
Field Theory.




\subsubsection{A remark on local observables}\label{sec:localobs}
Note that the model also possesses a local observable,
although its meaning is not transparent in the loop language.
Write $T_i=q^{H_i}$, so that
in the representation we use, $\pi_z(H_1)=-\pi_z(H_0)=\sigma^z$.
The coproduct of $H_i$ is the usual Lie algebra coproduct
$\Delta H_i = H_i\otimes 1 + 1\otimes H_i$,
so that the corresponding current
$h^{(\xsup)}=h_1^{(\xsup)}=-h_0^{(\xsup)}$ 
and $h^{(\tsup)}=h_1^{(\tsup)}=-h_0^{(\tsup)}$ does not carry a ``tail'', \ie, is local.
In the vertex language, 
$\aver{h^{(\xsup)}(x,t)}$ simply gives the average orientation
of the arrow sitting on the edge $(x,t)$, and similarly for $h^{(\tsup)}$.
This observable should not be confused with the flux observable associated
to $E_1$ or $\bar E_1$.

\subsection{Application to the dilute loop model}
\subsubsection{Loop observables associated to $E_0$ and $\bar E_0$}
Consider the dilute {\TL} loop model on the lattice $\Omega$ described in \secref{ssec:lattice},
with empty boundary conditions on all sides, except for one boundary site
which is pointing inwards. The reason for this choice is that
in the representation considered in~\secref{ssec:qaa}, the operator $E_0$ turns a spin $\u$ into $0$,
or a $0$ into $\d$. Therefore, in the loop model, inserting $E_0(x,t)$ forces
the path $\gamma$ to go from the boundary defect to $(x,t)$. 
For definiteness, we place 
the boundary defect at the bottom, say at $(x_d,1/2)$.
Consider firstly the case where the operator sits on a horizontal edge, $e_0^{(\tsup)}$:
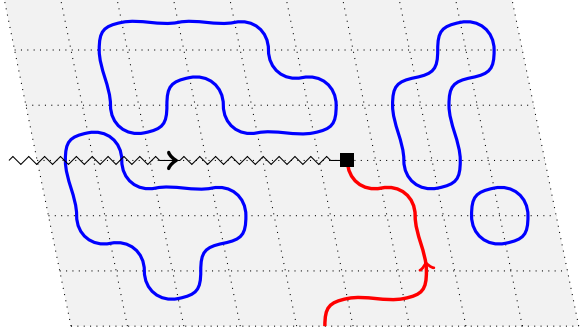
\begin{center}
\slow{  \begin{tikzpicture} [scale=0.75,distort]
    \fill[bgplaq,shift={(-0.5,-0.5)}] (0,0) rectangle (9,6);
    \draw[dotted,shift={(-0.5,-0.5)}] (0,0) grid (9,6);
    \draw[edge] (1,2.5) \start\south\go\east\go\east\go\south\go\west\go\south\go\west\go\north\go\west\go\north\go\north\go\east\go\south;
    \draw[edge] (2,3.5) \start\north\go\east\go\south\go\east\go\east\go\north\go\west\go\north\go\west\go\west\go\west\go\south\go\south\go\east\go\north;
    \draw[edge] (7,1.5) \start\south\go\east\go\north\go\west\go\south;
    \draw[edge] (6,2.5) \start\south\go\east\go\north\go\north\go\east\go\north\go\west\go\south\go\west\go\south\go\south;
    \draw[red,line width=1.3pt,
      decoration={markings, mark = at position 0.5 with {\arrow{>}}},
      postaction={decorate}]
    (4,-0.5) \start\north\go\east\go\east\go\north\go\north\go\west\go\north;
    \draw[wavy] (-1,2.5) -- (5,2.5) node[oper] {};
  \end{tikzpicture}}
\end{center}

For the open path $\gamma$, we get the following factors. Denote the total winding angle of $\gamma$
by $\theta$. It is an integer multiple of $\pi$: $\theta=k\pi$ (however we do not know anything
about the parity of $k$ this time because in the dilute model the $v$-tiles are parity-changing).
Then $e_0^{(\tsup)}$ gets a factor $e^{i\nu \theta}$ from the local turns.
Also, it is obvious that this loop crosses the tail (algebraically)
$\left\lfloor k/2 \right\rfloor$ times. This produces a factor of
$q^{2 \left\lfloor k/2 \right\rfloor}$ from the tail (since the tail is comprised of $T_0$ operators).
Also, the terms with $k$ odd get a factor $q$ from the matrix element of $E_0$. 
The final expression for the time-component of the current $e_0^{(\tsup)}$ is
\begin{equation*}
  \aver{e_0^{(\tsup)}(x,t)} = \frac{\varphi(q) w^{1-\ell}}{Z} \left(
  \sum_{\substack{C|\gamma:(x_d,1/2)\to(x,t) \\ \text{$k$ even}}} e^{i (\frac{\nu}{2} - \frac{1}{4}) k\pi} 
  + \sum_{\substack{C|\gamma:(x_d,1/2)\to(x,t) \\ \text{$k$ odd}}} q\ e^{i (\frac{\nu}{2} - \frac{1}{4}) (k-1)\pi} 
  \right) W(C) \ e^{i\nu k\pi} \,,
\end{equation*}
which can be combined into a single summation
\begin{equation*}
  \aver{e_0^{(\tsup)}(x,t)} = \frac{\varphi(q) w^{1-\ell}}{Z}
  \sum_{C|\gamma:(x_d,1/2)\to(x,t)}
  W(C) \ e^{i(\frac{3\nu}{2}-\frac{1}{4}) \theta} \,.
\end{equation*}

Let us move on to the case where the operator sits on a vertical edge, $e_0^{(\xsup)}$:
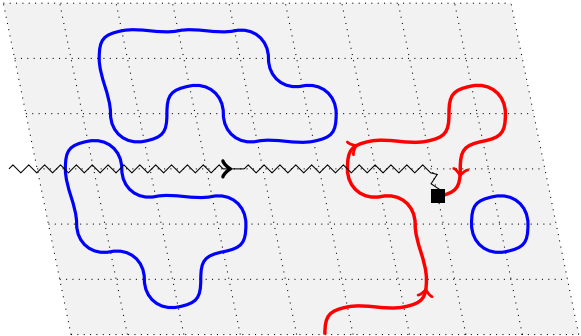
\begin{center}
\slow{
\begin{tikzpicture} [scale=0.75,distort]
    \fill[bgplaq,shift={(-0.5,-0.5)}] (0,0) rectangle (9,6);
\draw[dotted,shift={(-0.5,-0.5)}] (0,0) grid (9,6);
\draw[edge] (1,2.5) \start\south\go\east\go\east\go\south\go\west\go\south\go\west\go\north\go\west\go\north\go\north\go\east\go\south;
\draw[edge] (2,3.5) \start\north\go\east\go\south\go\east\go\east\go\north\go\west\go\north\go\west\go\west\go\west\go\south\go\south\go\east\go\north;
\draw[edge] (7,1.5) \start\south\go\east\go\north\go\west\go\south;
\draw[red,line width=1.3pt,
decoration={markings,
mark = at position 0.2 with {\arrow{>}}, 
mark = at position 0.5 with {\arrow{>}},
mark = at position 0.95 with {\arrow{>}}},
postaction={decorate}] (4,-0.5) \start\north\go\east\go\east\go\north\go\north\go\west\go\north\go\east\go\east\go\north\go\east\go\south\go\west\go\south\go\west;
    \draw[wavy] (-1,2.5) -- (6.5,2.5) -- (6.5,2) node[oper] {};
\end{tikzpicture}
}
\end{center}
We now have $\theta = k\pi-\alpha$ with $k \in \mathbb{Z}$, and as in the previous case,
the factor from the tail is $q^{2\left\lfloor k/2 \right\rfloor}$.
Putting everything together,
we can again combine the two parities of $k$ into a single summation
\begin{equation*}
  \aver{e_0^{(\xsup)}(x,t)} = \frac{\varphi(q) z^{1-\ell} e^{i(\frac{\nu}{2} - \frac{1}{4})\alpha}}{Z}
  \sum_{C|\gamma:(x_d,1/2)\to(x,t)}
  W(C) \ e^{i(\frac{3\nu}{2}-\frac{1}{4}) \theta} \,.
\end{equation*}
So far the value of $\ell$ (the constant introduced in the evaluation representation $\pi_z$) has been left arbitrary. In order to recover a discrete holomorphicity condition, 
we need the above formal expressions for $e_0^{(\xsup)}$ and $e_0^{(\tsup)}$ to differ
by a factor of $e^{i\alpha}$, and hence we require that
\begin{equation}
  \label{eq:fixingl}
  w^{1-\ell} = \left[z^{1-\ell} e^{i(\nu/2-1/4)\alpha} \right] \times e^{i\alpha} \,.
\end{equation}
Recalling that $(z/w)^{2\ell} = e^{-2i\nu\alpha}$,
one can check \eqref{eq:fixingl} is satisfied by choosing 
$\ell = \frac{2\nu}{3(2\nu+1)}$.
We now define
\begin{align*}
  \phi_0(x,t) &:= \varphi(q)^{-1} w^{\ell-1}
  \begin{cases}
    e_0^{(\tsup)}(x,t), & (x,t) \in (\mathbb{Z},\mathbb{Z} +\frac{1}{2}) \\
    e^{i\alpha} e_0^{(\xsup)}(x,t), & (x,t) \in (\mathbb{Z} +\half,\mathbb{Z}),
  \end{cases} \\
  \aver{\phi_0(x,t)} &= \frac{1}{Z} \sum_{C|\gamma:(x_d,1/2)\to(x,t)}
  W(C) \ e^{i(\frac{3\nu}{2}-\frac{1}{4}) \theta} \,.
\end{align*}
This observable satisfies the discrete holomorphicity condition:
\begin{equation}
  \phi_0(x,t-1/2)
  +e^{i(\pi-\alpha)}\phi_0(x+1/2,t)
  -e^{i(\pi-\alpha)}\phi_0(x-1/2,t)
  -\phi_0(x,t+1/2) = 0 \,.
\end{equation}
As in the dense case, choosing $\bar E_0$ instead of $E_0$ would lead to
the complex conjugate $\bar\phi_0$.

{\em Remark:} If one defines $\nu=\frac{1}{2}-\nu'$, then we get
$$
\aver{\phi_0(x,t)} = \frac{1}{Z} \sum_{C|\gamma:(x_d,1/2)\to(x,t)}
W(C) \ e^{-i(\frac{3\nu'}{2}-\frac{1}{2}) \theta} \,,
$$
and $x=z/w=e^{3i (\nu'-1)\alpha}$, which gives the same discretely holomorphic observable
as in~\cite{IkhlefC09}, but a different relationship between the ratio of spectral parameters
$x$ and the angle $\alpha$.

\subsubsection{Loop observables associated to $E_1$ and $\bar E_1$}
The operator $E_1$ takes a spin $\d$ to $\u$. Hence, as in the dense model,
we should consider two boundary defects, keeping the rest of the boundary
 empty for example.
The insertion of operator $E_1$ at a given point then 
selects loop configurations such that
the open path $\gamma$ that starts/ends at the boundary defects
passes through this point.
Let us now compute explicitly the resulting factors.
Firstly there is a factor of $w^{2\ell}$. For the winding of the $\gamma$ from the
boundary to the point in the bulk (and from the point in the bulk to the boundary),
we obtain a total factor of $e^{-2i\nu\theta}$, where $\theta = k\pi$. Finally,
for the crossings of the tail (which is now composed of $T_1$ matrices),
we obtain a factor of $q^{4k} = e^{i(2\nu-1)\theta}$. Putting everything together, we have
$$
\aver{e^{(\tsup)}_1(x,t)} = \frac{w^{2\ell}}{Z} \sum_{C | (x,t) \in \gamma} W(C) \ e^{-i\theta} \,.
$$
Notice that, in contrast to the dense case, here $e^{-i\theta}$ is not independent
of the configuration since the parity of $k$ is not fixed.

The calculation is completely analogous for $e_1^{(\xsup)}$. Firstly there is a factor of $z^{2\ell}$.
For the winding of $\gamma$ we obtain a total factor of $e^{-2i\nu\theta}$,
where $\theta = k\pi-\alpha$. Finally, for the crossings of the tail we obtain a
factor of $q^{4k} = e^{i(2\nu-1)(\theta+\alpha)}$. Putting everything together gives
$$
\aver{e^{(\xsup)}_1} = \frac{z^{2\ell} e^{i(2\nu-1)\alpha}}{Z}
\sum_{C | (x,t) \in \gamma} W(C) \ e^{-i\theta} \,.
$$
Using $w^{2\ell} = z^{2\ell} e^{2i\nu\alpha}$, we define a function on the edges of the lattice
\begin{align*}
  \phi_1(x,t) &:= w^{-2\ell}
  \begin{cases}
    e_1^{(\tsup)}(x,t), & (x,t)\in\mathbb{Z}\times(\mathbb{Z}+\half) \\
    e^{i\alpha} e_1^{(\xsup)}(x,t), & (x,t)\in(\mathbb{Z}+\half)\times\mathbb{Z},
  \end{cases} \\
  \aver{\phi_1(x,t)} &= \frac{1}{Z} \sum_{C | (x,t) \in \gamma} W(C) \ e^{-i\theta} \,,
\end{align*}
which satisfies the discrete holomorphicity condition
\begin{equation*}
  \phi_1(x,t-1/2)
  +e^{i(\pi-\alpha)}\phi_1(x+1/2,t)
  -e^{i(\pi-\alpha)}\phi_1(x-1/2,t)
  -\phi_1(x,t+1/2) = 0 \,.
\end{equation*}
Again, using $\bar E_1$ we obtain the antiholomorphic counterpart $\bar\phi_1$.

\section{Vertex models with integrable boundaries}
\label{sec:vertex-bound}
\subsection{Coideal subalgebras and \texorpdfstring{$K$}{K}-matrices}
Following the approach of~\cite{Skl88}, a  left (resp.\ right) boundary quantized affine algebra $B$ can be defined as a subalgebra of $U$ with the left (resp.\ right) coideal property $\Delta: B\ra B\ot U$ (resp.\ $\Delta: B \ra U \ot B$). 
If $V_z$ is irreducible as a $B$ module, then the left (resp.\ right) boundary reflection matrix 
$K_\ell(z): V_{z^{-1}}\ra V_{z}$ (resp.\ $K_r(z): V_{z} \ra V_{z^{-1}}$) is the solution, unique up to an overall normalization, of the linear relation 
\begin{equation} 
K_\ell(z) \pi_{z^{-1}}(Y)
=
\pi_{z}(Y) K_\ell(z)
\quad\quad
\Big(
\text{resp.}\ \ 
K_r(z) \pi_{z}(Y)
=
\pi_{z^{-1}}(Y) K_r(z)
\Big)
\label{eq:Kcomm}
\end{equation}
for all $Y \in B$.
We take the following as the graphical representation of $K_\ell(z)$ and $K_r(z)$:
\begin{equation*}
K_\ell(z)
=
    \begin{tikzpicture} [baseline=-3pt,scale=0.75]
    \path[left color=white,right color=lightgray] (0,-1) rectangle (-0.5,1); 
    \draw (0,-1) -- (0,1); 
    \draw[arr=0.5] (1,-1) node[right] {$z^{-1}$} -- (0,0); 
    \draw[arr=0.5] (0,0) -- 
    (1,1) node[right] {$z$} ; 
  \end{tikzpicture}
  \qquad
\text{and}
 \qquad
  K_r(z)
  =
   \begin{tikzpicture} [baseline=-3pt,scale=0.75]
\path[right color=white,left color=lightgray] (0,-1) rectangle (0.5,1); 
    \draw (0,-1) -- (0,1); 
\draw[arr=0.5] (-1,1) node[left] {$z$}-- (0,0);
\draw[arr=0.5] (0,0) -- (-1,-1) node[left] {$z^{-1}$};
\end{tikzpicture}
 \end{equation*}
where we recall that the arrows represent the flow of time as one reads
algebraic expressions from right to left.

>From this, Sklyanin~\cite{Skl88} defined a ``double-row transfer matrix'' $\mathbf{T}_2$:
 \[ 
\mathbf{T}_2(z;w_1,\dots,w_L)
= 
 {\rm Tr}_0
 \left(
 K_{{\rm r},0}(z) 
 R_{0L}(z/w_L) \dots R_{01}(z/w_1)  
 K_{{\rm l},0}(z) 
 R_{10}(zw_1) \dots   R_{L0}(zw_L)
 \right)
 \]
 which has the graphical representation
\begin{equation*}
\mathbf{T}_2(z;w_1,\dots,w_L)
= 
\begin{tikzpicture} [baseline=-3pt,scale=0.75]
\path[left color=white,right color=lightgray] (0,-1) rectangle (-0.5,1); 
\draw (0,-1) -- ++(0,2); 
\path[right color=white,left color=lightgray] (10,-1) rectangle (10.5,1); 
\draw (10,-1) -- ++(0,2); 
\draw[arr=0.05,rounded corners] (10,0) -- (9.5,-0.5) -- (0.5,-0.5) node[below] {$z^{-1}$} -- (0,0);
\draw[arr=0.95,rounded corners] (0,0) -- (0.5,0.5) node[above] {$z$} --(9.5,0.5) -- (10,0); 
\foreach\x in {1,2}
\draw[arr=0.2] (\x,-1.25) node[below] {$w_{\x}$}-- (\x,1.25);
\foreach\x in {3}
\draw[arr=0.2] (\x,-1.25) node[below] {$\cdots$}-- (\x,1.25);
\foreach\x in {4,...,8}
\draw[arr=0.2] (\x,-1.25) node[below] {$$}-- (\x,1.25);
\foreach\x in {9}
\draw[arr=0.2] (\x,-1.25) node[below] {$w_{L}$}-- (\x,1.25);
 \end{tikzpicture}
 \end{equation*}
By vertically stacking $M$ double-row transfer matrices $\mathbf{T}_2(z_i;w_1,\dots,w_L)$, $1\leq i \leq M$, one builds a rectangular lattice of width $L$ and height $2M$, whose left/right boundary conditions are fixed {\it a priori.} Each $\mathbf{T}_2$ is an operator acting in $V_{w_1} \otimes \cdots \otimes V_{w_L}$.

\subsection{Current conservation at the boundary}
\label{ssec:qg-bound}
We now extend the formalism of~\cite{BernardFelder91} to the case of an integrable boundary.
For simplicity, we treat only the case of the left boundary
in full detail, but analogous relations can also be written for right boundaries. 

Suppose we are given a left coideal subalgebra $B_\ell \subset U$, \ie,
a subalgebra with the additional property $\Delta(B_\ell)\subset B_\ell\otimes U.$\footnote{In view of the simple coproduct relations (\ref{eq:coprJ}--\ref{eq:coprth}), one way to produce such a $B_\ell$ is
to have it generated by appropriate combinations of the elements $J_a$ and $\thab{a}{b}$. We will discuss examples of such left coideal subalgebras in \secref{ssec:coideal}.} Assume also that we have a left boundary reflection matrix $K_\ell(z)$ which satisfies \eqref{eq:Kcomm}, for all $Y \in B_\ell$. Provided that $J_a \in B_\ell$, we have $K_\ell(z) \pi_{z^{-1}}(J_a)=\pi_{z}(J_a) K_\ell(z)$, which is represented graphically by
\begin{equation}\label{eq:leftK-J}
  \begin{tikzpicture} [baseline=-3pt,scale=0.75]
    \path[left color=white,right color=lightgray] (0,-1) rectangle (-0.5,1); 
    \draw (0,-1) -- (0,1); 
    \draw[arr=0.25] (1,-1)  node[right] {$z^{-1}$} -- (0,0);
    \draw[arr=0.5] (0,0)  -- (1,1) node[right] {$z$} ; 
    \draw[wavy] (-0.5,-0.5) node[left] {$a$} -- (0.5,-0.5) node[oper] {};
  \end{tikzpicture}
  \quad = \quad
  \begin{tikzpicture} [baseline=-3pt,scale=0.75]
    \path[left color=white,right color=lightgray] (0,-1) rectangle (-0.5,1); 
    \draw (0,-1) -- (0,1); 
    \draw[arr=0.5] (1,-1) node[right] {$z^{-1}$} -- (0,0); 
    \draw[arr=0.75] (0,0) -- (1,1) node[right] {$z$}; 
    \draw[wavy] (-0.5,0.4) node[left] {$a$} -- (0.5,0.4) node[oper] {};
  \end{tikzpicture}
  \end{equation}
We have included tails of operators, but since with our conventions they extend to the left of the operator insertion, they never cross any lines and therefore play no role in equation \eqref{eq:leftK-J}. 
This equation expresses the local conservation of the current at the left
boundary.

Similarly if $\thab{a}{b} \in B_\ell$, we have $K_\ell(z)\pi_{z^{-1}} (\thab{a}{b}) = \pi_{z}(\thab{a}{b}) K_\ell(z)$, which has the graphical representation  
\begin{equation}\label{eq:leftK-th}
  \begin{tikzpicture} [baseline=-3pt,scale=0.75]
    \path[left color=white,right color=lightgray] (0,-1) rectangle (-0.5,1); \draw (0,-1) -- (0,1); 
    \draw[arr=0.25] (1,-1) node[right] {$z^{-1}$} -- (0,0);
    \draw[arr=0.5] (0,0) -- (1,1) node[right] {$z$} ; 
    \draw[wavy=0.4] (0,-1) node[left] {$a$} -- (1,0) node[right] {$b$};
  \end{tikzpicture}
  \quad =\quad
\begin{tikzpicture} [baseline=-3pt,scale=0.75]
    \path[left color=white,right color=lightgray] (0,-1) rectangle (-0.5,1); \draw (0,-1) -- (0,1); 
    \draw[arr=0.5] (1,-1)  node[right] {$z^{-1}$} -- (0,0);
    \draw[arr=0.75] (0,0)  -- (1,1) node[right] {$z$} ; 
    \draw[wavy=0.4] (0,0.9) node[left] {$a$} -- (1,-0.1) node[right] {$b$};
  \end{tikzpicture}
 \end{equation}

Analogous relations apply if $J_a$ and $\thab{a}{b}$ are elements of a right coideal subalgebra $B_r$ with right boundary reflection matrix $K_r(z)$:
\begin{equation}\label{eq:rightK}
  \begin{tikzpicture} [baseline=-3pt,scale=0.75]
\path[right color=white,left color=lightgray] (0,-1) rectangle (0.5,1); 
    \draw (0,-1) -- (0,1); 
    \draw[arr=0.25] (-1,1) node[left] {$z$} -- (0,0); 
    \draw[arr=0.5] (0,0) -- (-1,-1) node[left] {$z^{-1}$}; 
    \draw[wavy] (-1.5,0.5) node[left] {$a$} -- (-0.5,0.5) node[oper] {};
  \end{tikzpicture}
  \quad = \quad
  \begin{tikzpicture} [baseline=-3pt,scale=0.75]
\path[right color=white,left color=lightgray] (0,-1) rectangle (0.5,1); 
    \draw (0,-1) -- (0,1); 
   \draw[arr=0.5] (-1,1) node[left] {$z$} -- (0,0); 
    \draw[arr=0.75] (0,0) -- (-1,-1) node[left] {$z^{-1}$}; 
     \draw[wavy] (-1.5,-0.5) node[left] {$a$} -- (-0.5,-0.5) node[oper] {};
  \end{tikzpicture}
\quad\quad\hbox{and}\quad\quad
  \begin{tikzpicture} [baseline=-3pt,scale=0.75]
\path[right color=white,left color=lightgray] (0,-1) rectangle (0.5,1); 
    \draw (0,-1) -- (0,1); 
    \draw[arr=0.25] (-1,1) node[left] {$z$} -- (0,0); 
    \draw[arr=0.5] (0,0) -- (-1,-1) node[left] {$z^{-1}$}; 
         \draw[wavy=0.4] (-1,0) node[left] {$a$} -- (0,1) node[right] {$b$};
  \end{tikzpicture}
  \quad = \quad
  \begin{tikzpicture} [baseline=-3pt,scale=0.75]
\path[right color=white,left color=lightgray] (0,-1) rectangle (0.5,1); 
    \draw (0,-1) -- (0,1); 
   \draw[arr=0.5] (-1,1) node[left] {$z$} -- (0,0); 
    \draw[arr=0.8] (0,0) -- (-1,-1) node[left] {$z^{-1}$}; 
 \draw[wavy=0.4] (-1,-0.1) node[left] {$a$} -- (0,-1.1) node[right] {$b$}; 
  \end{tikzpicture}
  \end{equation}

The result of the preceding equations is that we now obtain current and charge conservation laws which are \emph{exact,} rather than correct up to boundary terms, as they were in \secref{ssec:qg-bulk}. 
For example, current conservation \eqref{eq:conserv-j} is now automatic, since the assumption that the tail commutes with the left boundary is implicit in the condition $\thab{a}{b} \in B_\ell$. 

Similarly, by using equations \eqref{pic:RcoprJ} and (\ref{eq:leftK-J})--(\ref{eq:rightK}), we find that
\begin{equation*}
  \begin{tikzpicture} [baseline=-3pt,scale=0.75]
\path[left color=white,right color=lightgray] (0,-0.75) rectangle (-0.5,1.75); 
\draw (0,-0.75) -- ++(0,2.5); 
\path[right color=white,left color=lightgray] (9,-0.75) rectangle (9.5,1.75); 
\draw (9,-0.75) -- ++(0,2.5); 
    \draw[arr=0.05,rounded corners] (0,0.75) -- (0.5,1.25) -- (8.5,1.25) -- (9,0.75);
    \draw[arr=0.95,rounded corners] (9,0.75) -- (8.5,0.25) -- (0.5,0.25) -- (0,0.75);
    \draw[contour=0.75] 
    (0,-0.25) -- (9,-0.25); 
    \foreach\x in {1,...,8}
    \draw (\x,1.75) -- (\x,-0.75);
    \draw [wavy] (0,-0.4) node[left] {$a$} -- (5,-0.4) node[oper] {} ;
  \end{tikzpicture}
  \quad=\quad
  \begin{tikzpicture} [baseline=-3pt,scale=0.75]
\path[left color=white,right color=lightgray] (0,-0.75) rectangle (-0.5,1.75); 
\draw (0,-0.75) -- ++(0,2.5); 
\path[right color=white,left color=lightgray] (9,-0.75) rectangle (9.5,1.75); 
\draw (9,-0.75) -- ++(0,2.5); 
    \draw[arr=0.05,rounded corners] (0,0.25) -- (0.5,0.75) -- (8.5,0.75) -- (9,0.25);
    \draw[arr=0.95,rounded corners] (9,0.25) -- (8.5,-0.25) -- (0.5,-0.25) -- (0,0.25);
    \draw[contour=0.75] 
    (0,1.25) -- (9,1.25); 
    \foreach\x in {1,...,8}
    \draw (\x,-0.75) -- (\x,1.75);
    \draw [wavy] (0,1.1) node[left] {$a$} -- (5,1.1) node[oper] {} ;
  \end{tikzpicture}
\end{equation*}
Graphically, this says that any charge built from $J_a,\thab{a}{b} \in B_\ell,B_r$ commutes with the double-row transfer matrix; or equivalently, such a charge is conserved:
\begin{equation} 
\label{eq:conserv-Q-bd}
  \mathbf{J}_a(t-1/2) = \mathbf{J}_a(t+3/2) \,.
\end{equation}

\subsection{Light-cone lattice}\label{sec:lc}
Let us now consider a specialization of the parameters $w_i$ on the vertical lines in our lattice. 
 We assume $L$ to be even
and choose
\begin{equation*}
w_{j\ {\rm odd}}  = z, \quad w_{j\ {\rm even}} = z^{-1}, \quad 1 \leq j \leq L.
\end{equation*}
Assuming that $R(1) \propto P$, which is the case for both $U=U_q(A^{(1)}_1)$ and $U=U_q(A^{(2)}_2)$, this causes every second $R$-matrix present in the lattice to degenerate into a $P$-matrix. The resulting ``light-cone'' lattice has half the number of vertices, with the remaining ones being rotated by 45 degrees. The double-row transfer matrix becomes
\begin{equation*}
\mathbf{T}_2=
\begin{tikzpicture}[baseline=-3pt,scale=-0.9]
\draw[arr=0.05] (2,0.5) -- ++(-0.5,-0.5) -- ++(0.5,-0.5);
\foreach\x in {3,5,...,9}
{
\draw[arr=0.05] (\x+1,0.5) -- ++ (-1,-1);
\draw[arr=0.05] (\x,0.5) -- ++ (1,-1);
}
\foreach\x in {2,4,...,10}
{
\draw (\x+1,1.5) -- ++ (-1,-1);
\draw (\x,1.5) -- ++ (1,-1);
}
\draw[arr=0.05] (11,0.5) -- ++ (0.5,-0.5) -- ++(-0.5,-0.5);
\path[left color=white,right color=lightgray] (11.5,-0.5) rectangle ++ (0.5,2); \draw (11.5,-0.5) -- (11.5,1.5);
\path[right color=white,left color=lightgray] (1,-0.5) rectangle ++ (0.5,2); \draw (1.5,-0.5) -- (1.5,1.5); 
\end{tikzpicture}
\end{equation*}
where we have absorbed an $R$-matrix in the right boundary;
denote the resulting boundary operator $\tilde K_r(z)$ (we skip
the details since we shall focus on the left boundary in what follows).
All lines are now oriented upwards, so that we omit orientation arrows henceforth. Equivalently, $\mathbf{T}_2 =\mathbf{T}_e \mathbf{T}_o$, where
\begin{align*}
\mathbf{T}_{\rm e} := K_\ell(z)
\prod_{i} \check R_{2i}(z^2) \tilde K_r(z) &=
\begin{tikzpicture}[baseline=-3pt,scale=0.9]
\useasboundingbox (0,-0.5) -- (12,0.5);
\foreach\x in {2,4,...,8}
{
\draw (\x,-0.5) -- ++ (1,1);
\draw (\x,0.5) -- ++ (1,-1);
}
\draw (1,0.5) -- ++(-0.5,-0.5) -- ++(0.5,-0.5);
\draw (10,-0.5) -- ++(0.5,0.5) -- ++(-0.5,0.5);
\path[right color=white,left color=lightgray] (10.5,-0.5) rectangle ++ (0.5,1); \draw (10.5,-0.5) -- ++(0,1);
\path[left color=white,right color=lightgray] (0,-0.5) rectangle ++ (0.5,1); \draw (0.5,-0.5) -- ++(0,1); 
\end{tikzpicture}
\\
\mathbf{T}_{\rm o} := \prod_{i} \check R_{2i+1}(z^2)&=
\begin{tikzpicture}[baseline=-3pt,scale=0.9]
\useasboundingbox (0,-0.5) -- (12,0.5);
\foreach\x in {1,3,...,9}
{
\draw (\x,-0.5) -- ++ (1,1);
\draw (\x,0.5) -- ++ (1,-1);
}
\end{tikzpicture}
\end{align*}
$\check R_i=P_{i,i+1}R_{i,i+1}$ is the $R$-matrix acting
on sites $i,i+1$ with an additional permutation $P$ of factors of the tensor
product, and $K_\ell(z)$ (resp.\ $\tilde K_r(z)$) acts on the first (resp.\ last) factor of the tensor product.

Since the light-cone lattice is obtained as a special case of the double-row, two-boundary lattice, all previous results continue to apply; they only need to be transcribed into the new orientation. The analogue of relation \eqref{pic:RcoprJ} is the commutation of $\check R$ with the action of $J_a$, that is:
\begin{equation}
\label{eq:local_conserv_lc}
\begin{tikzpicture}[baseline=-3pt,scale=0.9]
\draw (-1,-1) -- ++ (2,2);
\draw (1,-1) -- ++ (-2,2);
\draw[wavy=0.45] (-2,-0.5) node[below] {$a$} -- (-0.5,-0.5) node[oper] {};
\end{tikzpicture}
+
\begin{tikzpicture}[baseline=-3pt,scale=0.9]
\draw (-1,-1) -- ++ (2,2);
\draw (1,-1) -- ++ (-2,2);
\draw[wavy] (-2,-0.5) node[below] {$a$} -- (0.5,-0.5) node[oper] {};
\end{tikzpicture}
=
\begin{tikzpicture}[baseline=-3pt,scale=0.9]
\draw (-1,-1) -- ++ (2,2);
\draw (1,-1) -- ++ (-2,2);
\draw[wavy=0.45] (-2,0.5) node[below] {$a$} -- (-0.5,0.5) node[oper] {};
\end{tikzpicture}
+
\begin{tikzpicture}[baseline=-3pt,scale=0.9]
\draw (-1,-1) -- ++ (2,2);
\draw (1,-1) -- ++ (-2,2);
\draw[wavy] (-2,0.5) node[below] {$a$} -- (0.5,0.5) node[oper] {};
\end{tikzpicture}
\end{equation}
The 45 degree rotation makes ``space'' (resp.\ ``time'')
lines go south-west to north-east (resp.\ south-east to north-west).
The corresponding currents are simply
\begin{eqnarray*}
j_a(x)&=&
\begin{tikzpicture}[baseline=-3pt,scale=0.90]
\foreach\x in {1,3,...,9}
\draw (\x,-0.25) -- ++ (0.5,0.5) ++ (0.5,0) -- ++ (0.5,-0.5);
\draw (7,-0.25) -- ++ (0.5,0.5);
\draw[wavy=0.4] (0.5,0) node[below] {$a$} -- (4.25,0) node[oper] {} node[below=1mm] {$\ss x$};
\end{tikzpicture}\\\wh{\jmath}_a(x)&=&
\hspace*{7mm}\begin{tikzpicture}[baseline=-3pt,scale=0.90]
\foreach\x in {1,3,...,9}
\draw (\x,-0.25) -- ++ (0.5,0.5) ++ (0.5,0) -- ++ (0.5,-0.5);
\draw (7,-0.25) -- ++ (0.5,0.5);
\draw[wavy=0.4] (11,0) node[below] {$a$} --(4.25,0) node[oper] {} node[below=1mm] {$\ss x$}  ;
\end{tikzpicture}\\
\end{eqnarray*}
(or opposite tilt of the lines depending on parity of $t$), 
where $x\in\mathbb{Z}$ and
the former distinction between ``time'' and ``space'' components
is unnecessary.
The same conservation equation \eqref{eq:conserv-j} 
holds as a consequence of
\eqref{eq:local_conserv_lc}, and the fact that $\thab{a}{b} \in B_\ell$. 
Rewritten in the light-cone approach, it becomes:
\begin{equation}\label{eq:conserv-jj}
 j_a(x,t) + j_a(x+1,t)
 = j_a(x,t+1) + j_a(x+1,t+1)
\,,\qquad (x,t)\in\mathbb{Z}^2,\ x+t=0\pmod{2}\,.
\end{equation}

Similarly, \eqref{eq:leftK-J} can be rewritten algebraically as
\begin{equation}\label{eq:leftconserv}
j_a(1,t)=j_a(1,t+1)\,,\qquad t=0\pmod{2}\,.
\end{equation}
There is a right boundary analogue:
\begin{equation}\label{eq:rightconserv}
j_a(L,t)=j_a(L,t+1)\,,\qquad t=0\pmod{2}\,.
\end{equation}

The charge $\mathbf{J}_a$ is now the obvious sum: 
$\mathbf{J}_a=\sum_{x=1}^L j_a(x)$,
or graphically,
\begin{equation}
\label{eq:chargelc}
\mathbf{J}_a
=
\begin{tikzpicture}[baseline=-3pt,scale=0.90]
\foreach\x in {1,3,...,9}
\draw (\x,-0.25) -- ++ (0.5,0.5) ++ (0.5,0) -- ++ (0.5,-0.5);
\draw (7,-0.25) -- ++ (0.5,0.5);
\draw[wavy] (0.5,0.075) node[below] {$a$} -- (5.25,0.075) node[oper] {};
\draw[contour] (0.5,-0.075) -- (11,-0.075);
\end{tikzpicture}
\end{equation}
and as a direct consequence of 
(\ref{eq:conserv-jj}--\ref{eq:rightconserv}),
commutes with both $\mathbf{T}_{\rm e}$ and $\mathbf{T}_{\rm o}$, \ie,
\begin{equation}\label{eq:globalconserv}
\mathbf{J}_a(t)=\mathbf{J}_a(t+1)\,.
\end{equation}

As already pointed out, we are mainly concerned with local current conservation
rather than conservation of the charge;
in particular the local conservation at the left boundary \eqref{eq:leftconserv} only requires integrability at the left boundary.

\subsection{Adjoint action on the light-cone lattice}\label{sec:adj-lc}
The adjoint action of an element
$J_a$ of $U$ on another element $J_b$ is defined as in~\eqref{eq:adj-graph}.
On the light-cone lattice, and in the 
case of an operator $J_a$ commuting with $K$-matrices
on all boundaries, the expression of
$A_a[j_b^{(\tsup)}]$ can be greatly simplified. We illustrate this by a series of pictures in which,
in preparation for the switch to loop models, we use dual lattice pictures,
\cf \eqref{eq:plaq} in the bulk, and at the boundary, 
$
K_\ell = 
\begin{tikzpicture}[baseline=-3pt,scale=0.75,rotate=\lcrot,distort]
\bplaqw(0,0)
\end{tikzpicture}
$
and similarly for the other boundaries.

Using the conservation equation \eqref{eq:globalconserv},
the contour defining $A_a[j_b^{(\tsup)}]$ can be deformed to follow the top
and bottom boundaries, \ie,
\begin{align*}
A_a[j_b(x,t)]&=
\begin{tikzpicture}[baseline=0,rotate=\lcrot,distort]%
\newcount\u\newcount\v
\foreach\x in {-5,...,5}
\foreach\y in {-5,...,5}
{
  \pgfmathsetcount{\u}{\x+\y}
  \pgfmathsetcount{\v}{\x-\y}
\pgfmathrandominteger{\rand}{0}{1}
\ifnum\u<4
  \ifnum\u>-4
     \ifnum\v<7
        \ifnum\v>-7
        \plaq(\x,\y)
        \else
        \fi
     \else
     \fi
   \else
   \fi
\else
\fi
\ifnum\u=4
  \bplaqn(\x,\y)
\else
  \ifnum\u=-4
     \ifnum\y=-2
     \else
        \bplaqs(\x,\y)
     \fi
  \else
  \fi
\fi
\ifnum\v=7
   \bplaqe(\x,\y)
\else
  \ifnum\v=-7
     \bplaqw(\x,\y)
  \else
  \fi
\fi
}
\draw[wavy] (-3.5,4) node[left] {$b$} -- (0,0.5) node[oper] {};
\draw[contour] (-3.4,4.1) -- (4.1,-3.4) -- (3.85,-3.65)-- (-3.65,3.85) ;
\draw[wavy] (-3.35,4.2) node[left] {$a$} -- (1.35,-0.5) node[oper] {};
\end{tikzpicture}
\\
&=
\begin{tikzpicture}[baseline=0,rotate=\lcrot,distort] %
\newcount\u\newcount\v
\foreach\x in {-5,...,5}
\foreach\y in {-5,...,5}
{
  \pgfmathsetcount{\u}{\x+\y}
  \pgfmathsetcount{\v}{\x-\y}
\pgfmathrandominteger{\rand}{0}{1}
\ifnum\u<4
  \ifnum\u>-4
     \ifnum\v<7
        \ifnum\v>-7
        \plaq(\x,\y)
        \else
        \fi
     \else
     \fi
   \else
   \fi
\else
\fi
\ifnum\u=4
  \bplaqn(\x,\y)
\else
  \ifnum\u=-4
     \ifnum\y=-2
     \else
        \bplaqs(\x,\y)
     \fi
  \else
  \fi
\fi
\ifnum\v=7
   \bplaqe(\x,\y)
\else
  \ifnum\v=-7
     \bplaqw(\x,\y)
  \else
  \fi
\fi
}
\draw[wavy] (-3.5,4) node[left] {$b$} -- (0,0.5) node[oper] {};
\draw[contour] (-2,5.5) -- (5.5,-2) -- (2,-5.5) -- (-5.5,2) ;
\draw[wavy] (-1.95,5.6) node[above] {$a$} -- (2.1,1.55) node[oper] {};
\end{tikzpicture}
\end{align*}
We now suppose that top and bottom boundary conditions are also integrable, such that $J_a$
commutes with the corresponding $K$-matrices, except
at some boundary defects which sit say on the bottom
boundary. Then the top part of the contour can be moved through the
top row of $K$-matrices, using again the boundary conservation relations. Indeed, the sum over the two points on each triangle along the top boundary gives a zero contribution.\footnote{To see this, we must pay attention to a crucial sign issue -- our currents are associated to edges
which are oriented upwards, so current conservation at the top/bottom
boundaries must be accompanied by a sign in one of the terms. It is precisely this sign which causes the pairwise cancellation for each triangle.} On the bottom part, the contour is reduced
to an arch enclosing the defects:
\begin{align*}
A_a[j_b(x,t)]&=
\begin{tikzpicture}[baseline=0,rotate=\lcrot,distort] %
\newcount\u\newcount\v
\foreach\x in {-5,...,5}
\foreach\y in {-5,...,5}
{
  \pgfmathsetcount{\u}{\x+\y}
  \pgfmathsetcount{\v}{\x-\y}
\pgfmathrandominteger{\rand}{0}{1}
\ifnum\u<4
  \ifnum\u>-4
     \ifnum\v<7
        \ifnum\v>-7
        \plaq(\x,\y)
        \else
        \fi
     \else
     \fi
   \else
   \fi
\else
\fi
\ifnum\u=4
  \bplaqn(\x,\y)
\else
  \ifnum\u=-4
     \ifnum\y=-2
     \else
        \bplaqs(\x,\y)
     \fi
  \else
  \fi
\fi
\ifnum\v=7
   \bplaqe(\x,\y)
\else
  \ifnum\v=-7
     \bplaqw(\x,\y)
  \else
  \fi
\fi
}
\draw[wavy] (-3.5,4) node[left] {$b$} -- (0,0.5) node[oper] {};
\draw[contour] (-2,-2.5) -- (-1.5,-2) -- (-2,-1.5)-- (-2.5,-2);
\draw[wavy] (2,-5.5) node[right] {$a$} -- (-1.5,-2) node[oper] {};
\end{tikzpicture}
\end{align*}
Translating this final picture into algebraic form, we obtain
\begin{equation} \label{eq:adjoint-defect}
\aver{  A_a[j_b(x,t)] }=
\aver{\left[\wh{\jmath}_a(x_d,0) + \wh{\jmath}_a(x_d+1,0) \right] j_b(x,t)} \,,
\end{equation}
where we assume that there are two defects at adjacent locations $x_d$ and $x_d+1$, and we recall that $\wh{\jmath}_a$ denotes a non-local current with a ``right'' tail.

\subsection{Coideal subalgebras for quantized affine algebras}
\label{ssec:coideal}
Here we give examples of (left) coideal subalgebras and boundary reflection matrices for the quantized affine algebras that interest us. Since we do not discuss analogous right boundary results, we omit the subscript ``$\ell$'' from all subsequent equations. 

\subsubsection{The $ U_q(A_1^{(1)})$  boundary algebra}
\label{sssec:bound-alg-1}
The choice of coideal subalgebra $B$ 
we shall consider in this paper is generated by
\begin{equation*}
  \{
  T_0 \,,  T_1 \,,
  Q:= E_1 + r \bar E_0 \,, 
  \bar Q:=\bar E_1 + r E_0
  \} \,,
\end{equation*}
where $r$ is a real parameter. The left coideal property is satisfied because
\begin{equation*}
\Delta(Q) = 
Q \otimes 1 + T_1 \otimes E_1 + T_0 \otimes r \bar E_0
\qquad \text{and} \qquad
\Delta(\bar Q)=
\bar Q \otimes 1 + T_1 \otimes \bar E_1 + T_0 \otimes r E_0
\end{equation*}
are both elements of $B \otimes U$.
After a choice of normalization the solution of (\ref{eq:Kcomm}) is~\cite{NepMez98}
\begin{equation} \label{eq:K-6V}
  K(z) =\left( \begin{array}{cc}
    z+rz^{-1} & 0 \\
    0 & z^{-1} + rz
  \end{array} \right) \,.
\end{equation}
Note that $K(z)$ is diagonal as a consequence of the fact that $T_0$ and $T_1$ are elements of $B$.

\subsubsection{The $ U_q(A_2^{(2)})$  boundary algebra}
\label{sssec:bound-alg-2}
The boundary algebra $B$ in this case is taken to be that generated by 
$\{T_0,T_1,E_1,\bar E_1,Q, \bar Q\}$, where
\begin{align*}
  Q := [E_1,E_0]_{q^{-4}} + r \bar E_0 \,,
  \qquad 
  \bar Q := [\bar E_1,\bar E_0]_{q^4} -r q^2 E_0 \,.
\end{align*}
in which we use the notation $[a,b]_x=ab-xba$. 

The left coideal property $\Delta(B)\in B\ot U$ is easy
to check. With a choice of normalization, the solution of~\eqref{eq:Kcomm} in the case when $r$ is fixed to be 
$r=\pm i q^{-1}$ reads~\cite{BFKZ96,Nep02} 
\begin{equation} \label{eq:K-19V}
  K(z) = \left( \begin{array}{ccc}
    z^{2\ell}(z^{-1}+ r z) & 0 & 0 \\
    0 & z+ r z^{-1} & 0 \\
    0 & 0 & z^{-2\ell}(z^{-1}+r z)
  \end{array} \right) \,.
\end{equation}
For definiteness, we shall henceforth take the root $r=iq^{-1}$. As in the $U_q(A_1^{(1)})$ model case , $K(z)$ is diagonal because $T_0$ and $T_1$ are elements of $B$. In contrast to the $U_q(A_1^{(1)})$ case, there is no solution of \eqref{eq:Kcomm} for general values of the parameter $r$.

\section{Integrable boundaries for loop models}
\label{sec:loop-bound}
In this section we repeat the ideas of \secref{sec:vertex-loop} to introduce boundary tiles into the dense and dilute \TL\ models. In complete analogy with \secref{sec:vertex-loop}, the corresponding $K$ matrices \eqref{eq:K-6V} and \eqref{eq:K-19V} are recovered as linear combinations of the boundary tiles.
We use the light-cone approach of \secref{sec:lc}, with an angle of $\alpha$ on the lattice, \ie,
\begin{align*}
\cR &=
\begin{tikzpicture}[baseline=-3pt,scale=0.75,rotate=\lcrot,distort]
\draw (1,0) -- (-1,0) node[below] {$z$};
\draw (0,1) -- (0,-1) node[below] {$z^{-1}$};
\draw (-0.2,0) arc (180:90:0.2cm) node[left] {$\alpha$};
\end{tikzpicture}
=\ 
\begin{tikzpicture}[baseline=-3pt,scale=1.25,rotate=\lcrot,distort]
\plaq(0,0)
\end{tikzpicture}
\\
K &=
\begin{tikzpicture}[baseline=-3pt,scale=0.75]
\path[left color=white,right color=lightgray] (0,-0.75) rectangle (-0.5,0.75); 
\draw (0,-0.75) -- (0,0.75); 
\begin{scope}[rotate=\lcrot,distort]
    \draw (0,-1) node[right] {$z^{-1}$} -- (0,0); 
    \draw (0,0) -- 
    (1,0) node[right] {$z$} ; 
\draw (0.2,0) arc (0:-90:0.2cm) node[right] {$\alpha$};
  \end{scope}
\end{tikzpicture}
=\ 
\begin{tikzpicture}[baseline=-3pt,scale=1.25,rotate=\lcrot,distort]
\bplaqw(0,0)
\end{tikzpicture}
\end{align*}
and similarly for other boundaries.

\subsection{The boundary dense \TL{} model and the 
\texorpdfstring{$U_q(A_1^{(1)})$}{Uq(A1(1))}
vertex model}
Let us define a boundary loop model by introducing an additional weight 1 boundary plaquette, as follows:
$$
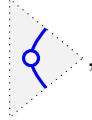
\begin{tikzpicture} [baseline=-3pt,scale=1.25,rotate=\lcrot,distort]
\bplaqwb(0,0)
\end{tikzpicture}
\,,
$$
and by assigning weight $\tau^{(n)}=-(e^{i\nu(2\pi - n\xi)}+ e^{- i\nu (2\pi - n\xi)})$ to any loop that that passes 
$n$ times through a boundary. Thus we can view the boundary as introducing a deficit angle of $\xi$.

Again, we can turn this boundary weight into that of a vertex model by viewing it as an operator from $V\to V$ in 
a S-N direction, resulting in a boundary weight
\begin{equation}
\label{eq:newK-6V} 
K= \begin{pmatrix} e^{- i\nu(\alpha-\xi)} & 0\\ 0& e^{i\nu(\alpha-\xi)} 
\end{pmatrix} \,.
\end{equation}
Since the spectral parameter $z$ is related to the angle $\alpha$ by $z=e^{-i\nu\alpha}$, this boundary matrix coincides with $K(z)$ of ~\eqref{eq:K-6V}\footnote{In fact the two $K$-matrices coincide after renormalizing \eqref{eq:newK-6V} by 
$((1+rz^{-2})(1+rz^2))^{1/2}$. Since this only produces a global factor in the observables we will consider, we omit this normalization for simplicity.} if we take
\[ e^{2i\nu\xi}=(1+rz^{-2})/(1+rz^2).\]
Clearly the $r=0$ case corresponds to a zero deficit angle $\xi=0$ in which case the boundary TL plaquette becomes the 
single plaquette with weight one which we call ``free boundary conditions'' and denote by
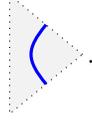
\[
\begin{tikzpicture} [baseline=-3pt,scale=1.25,rotate=\lcrot,distort]
\bplaqwa(0,0)
\end{tikzpicture}
\,.
\]

\subsection{The boundary dilute \TL{} model and the
\texorpdfstring{$U_q(A_2^{(2)})$}{Uq(A2(2))}
vertex model}
We introduce two additional boundary plaquette configurations with associated weights $\rho$ and $\kappa$, as follows:
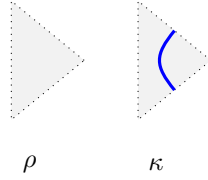
\[
    \begin{tikzpicture} [baseline=-3pt,scale=1.25,rotate=\lcrot,distort]
      \bplaqwz(0,0)
      \node at (-0.75,-1) {$\rho$};
    \end{tikzpicture}\qquad  
    \begin{tikzpicture} [baseline=-3pt,scale=1.25,rotate=\lcrot,distort]
      \bplaqwa(0,0)
      \node at (-0.75,-1) {$\kappa$};
    \end{tikzpicture}
\]
Interpreting the plaquettes as operators as above yields a boundary reflection matrix        
\begin{equation} 
K=\begin{pmatrix} \kappa e^{-i\nu \alpha} & 0 &0\\ 0 &\rho & 0\\ 0& 0& \kappa e^{i\nu \alpha} \end{pmatrix},
\end{equation}
which is equal to the matrix $K(z)$ of equation ~\eqref{eq:K-19V} if $\rho=z+iq^{-1}z^{-1}$ and $\kappa=z^{-1}+iq^{-1}z$, using also the fact that $z^{2\ell} = e^{-i\nu\alpha}$.

\section{Non-local currents and boundary discrete holomorphicity}
\label{sec:DH-bound}
\subsection{Application to the dense loop model}
\label{ssec:bound-dense}
For convenience, in this section we shall use exclusively the light-cone orientation of the lattice. We consider loop configurations on the lattice which contain a single open path $\gamma$. For simplicity, we assume that the ends of $\gamma$ are situated next to each other, as follows:  
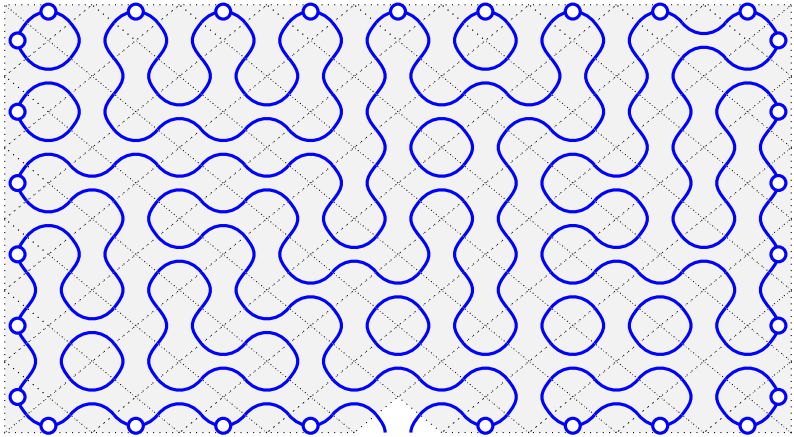
\begin{center}
\slow{
\begin{tikzpicture} [scale=0.75,rotate=\lcrot,distort]
\newcount\u\newcount\v
\pgfmathsetseed{45}
\foreach\x in {-7,...,7}
\foreach\y in {-7,...,7}
{\pgfmathrandominteger{\rand}{0}{1}
\pgfmathsetcount{\u}{\x+\y}
\pgfmathsetcount{\v}{\x-\y}
\ifnum\u<6
  \ifnum\u>-6
     \ifnum\v<9
        \ifnum\v>-9
           \ifnum\rand=0\plaqb(\x,\y)\else\plaqa(\x,\y)\fi
        \else
        \fi
     \else
     \fi
   \else
   \fi
\else
\fi
\ifnum\u=6
  \bplaqnb(\x,\y)
\else
  \ifnum\u=-6
     \ifnum\y=-3
     \else
        \bplaqsb(\x,\y)
     \fi
  \else
  \fi
\fi
\ifnum\v=9
   \bplaqeb(\x,\y)
\else
  \ifnum\v=-9
     \bplaqwb(\x,\y)
  \else
  \fi
\fi
}
\begin{scope}[shift={(-3,-3)}]
\clip (-0.5,0.5) -- (0.5,-0.5) -- (1,0) -- (0,1) -- cycle;
\draw[edge] (0.5,0) \start\west\go\south;
\draw[edge] (0,0.5) \start\south\go\west;
\end{scope}
\end{tikzpicture}
}
\end{center}
As we did in the case of trivial boundary conditions, we consider observables which are constructed by requiring that the open loop goes through a certain point $(x,t)$, either in the bulk or on the boundary of the lattice. Because the $K$-matrices are diagonal the correct way to obtain such observables is, as before, to insert a local operator (say $E_0$, complete with its tail) at $(x,t)$, since all lattice configurations vanish for which $E_0$ is situated on a closed loop: 
\begin{equation*}
\begin{tikzpicture}[baseline=-3pt,scale=0.75]
\draw[edge] (0,0) arc (0:360:0.8cm) node[oper,label={left:$E_0$}] {}
node[blob,pos=0.38] {}
node[blob,pos=0.62] {}
node[pos=0.473,right] {$\vdots$};
\end{tikzpicture}
\ =0 \,.
\end{equation*}

\subsubsection{Loop observables in the bulk}
Let us repeat the analysis of the observables in the bulk,
but now in the presence of non-trivial boundaries and on the light-cone lattice. 
Insert $e_0$ at the point $(x+1,t) \in \mathbb{Z}^2$, $x+t=0\ \text{(mod 2)}$, in the
bulk of the lattice. As mentioned above, the only configurations which survive are those for
which the open loop goes through $(x+1,t)$. The contribution of the open path to the
weight can be found in a similar way as before as $e^{i\nu(2\theta-\pi+n\xi)}q^k$,
where $\theta=k\pi$ is the angle formed by the left-incoming portion of the open loop 
(if we treat the boundary tiles on equal footing with bulk tiles),
and $n$ the number of contacts of the left portion of this loop with the boundary minus
that of the right portion. So we find
\begin{equation*}
  \aver{e_0(x+1,t)} = \frac{z^{-1}e^{-i\nu\pi}}{Z}
  \sum_{C|(x+1,t)\in \gamma} W(C) \ e^{i(4\nu-1)\theta} \ e^{ni\nu\xi} \,,
\end{equation*}
Similarly, repeating for $e_0(x,t)$ with $x+t=0\ \text{(mod 2)}$, we obtain
\begin{equation*}
  \aver{e_0(x,t)} = \frac{z\,e^{-i\nu\pi}e^{i(2\nu-1)\alpha}}{Z}
  \sum_{C|(x,t)\in \gamma} W(C) \ e^{i(4\nu-1)\theta} e^{ni\nu\xi} \,.
\end{equation*}

The local conservation law \eqref{eq:conserv-jj} with $j_a = e_0$ is
\begin{equation}\label{eq:conserv-ee}
e_0(x,t) + e_0(x+1,t)
=
e_0(x,t+1) + e_0(x+1,t+1)
\,,\qquad (x,t)\in\mathbb{Z}^2,\ x+t=0\pmod{2} \,.
\end{equation}
This holds in the bulk because the tail, which is comprised of $T_0$ operators, commutes with the left $K$-matrix (since $T_0 \in B$, as explained in \secref{sssec:bound-alg-1}). Therefore, using analogous arguments to those of \secref{ssec:loop-dense}, we define the function
\begin{equation*}
  \phi_0(x,t) := z e^{i\nu\pi}
  \begin{cases}
    e_0(x,t), & x+t=1\pmod 2 \\ 
    e^{i\alpha} e_0(x,t), & x+t=0\pmod 2\,, 
  \end{cases}
\end{equation*}
and in view of the fact that $z=e^{-i\nu\alpha}$, we have 
\begin{equation} \label{eq:phi0_bdry}
  \aver{\phi_0(x,t)} = \frac{1}{Z}
  \sum_{C|(x,t)\in \gamma} W(C) \ e^{i(4\nu-1)\theta} e^{ni\nu\xi}\,.
\end{equation}
Applying \eqref{eq:conserv-ee} to this observable, we find that it satisfies
\begin{equation*}
e^{i(\pi-\alpha)} \phi_0(x,t)
+
\phi_0(x+1,t)
-
e^{i(\pi-\alpha)} \phi_0(x+1,t+1)
-
\phi_0(x,t+1)
=
0\,,
\end{equation*} 
which is discrete holomorphicity on the light-cone lattice.

The observable corresponding to insertion of $E_1$ gets modified in an analogous way; namely
\begin{equation*}
  \phi_1(x,t) := z e^{-i\nu\pi}
  \begin{cases}
    e_1(x,t), & x+t=1\pmod 2 \\ 
    e^{i\alpha} e_1(x,t), & x+t=0\pmod 2\, 
  \end{cases}
\end{equation*}
gives rise to the observable 
\begin{equation} \label{eq:phi1_bdry}
  \aver{\phi_1(x,t)} = \frac{1}{Z}
\sum_{C|(x,t)\in \gamma} W(C) \ e^{-i\theta} \ e^{-ni\nu\xi}\,.
\end{equation}
This is the flux observable in the presence of a non-trivial boundary and it satisfies the discrete holomorphicity equation
\begin{equation*}
e^{i(\pi-\alpha)} \phi_1(x,t)
+
\phi_1(x+1,t)
-
e^{i(\pi-\alpha)} \phi_1(x+1,t+1)
-
\phi_1(x,t+1)
=
0\,.
\end{equation*}

\subsubsection{Loop observables at the boundary}
By \emph{boundary} we mean in what follows the {\em left}\/ boundary. 
Since $r\neq 0$,
the operators $E_0,E_1,\bar E_0,\bar E_1$ are not elements of the 
coideal $B$,
and hence they are not conserved separately at the boundary (with trivial
boundaries, $E_1$ and $\bar E_1$ were conserved), 
and in particular
their associated charge will not be conserved.

We now consider 
the combinations $Q=E_1+r\bar E_0$ and $\bar Q=\bar E_1+r E_0$, which are
in $B$. Using \eqref{eq:leftconserv} in the case $j_a=e_1+r\bar e_0$ we find that
\begin{equation*}
e_1(1,t) + r \bar e_0(1,t)
=
e_1(1,t+1) + r \bar e_0(1,t+1)\,,
\qquad t=0\pmod{2} \,,
\end{equation*}
which can be translated into the following equation for the observables
$\phi_1$ and $\bar\phi_0$:
\begin{equation*}
z^{-1} \phi_1(1,t) + rz \bar\phi_0(1,t)
=
e^{-i\alpha} z^{-1} \phi_1(1,t+1) + e^{i\alpha} rz \bar\phi_0(1,t+1) \,.
\end{equation*}
This is neither a holomorphicity nor an antiholomorphicity condition
because we are mixing operators from the two chiralities. However, by taking the real part (or equivalently, summing this identity and the one
satisfied by the conjugate observable $\bar Q$),
one finds that $\psi:=z^{-1}(\phi_1+r\phi_0)$ satisfies
\begin{equation*}
  \mathrm{Re} \left[\psi(1,t)+e^{i(\pi-\alpha)}\psi(1,t+1) \right]=0 \,,
\end{equation*}
which is a boundary discrete holomorphicity condition around the plaquette
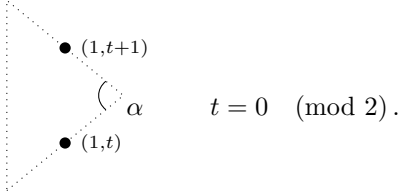
\begin{equation*}
  \label{eq:bound-plaq}
  \begin{tikzpicture} [rotate=\lcrot,distort,baseline=1cm]
  \draw[dotted] (0,0) -- (2,0) -- (2,2) -- cycle;
  \draw (2,0.3) arc (90:180:0.3) node[right=1.5mm] {$\alpha$};
  \node[circle,fill,inner sep=1.5pt,label={right:$\ss(1,t)$}] at (1,0) {};
  \node[circle,fill,inner sep=1.5pt,label={right:$\ss(1,t+1)$}] at (2,1) {};
  \end{tikzpicture}
  \qquad t=0\pmod 2 \,.
\end{equation*}
{\em Remark:} At the left boundary the tails (which are on the left) disappear
and therefore the two observables $\phi_1$ and $\bar\phi_0$ are 
the same up to a constant. This can be seen more explicitly in the fact that
there cannot be any winding at the boundary, so the angle $\theta$
in (\ref{eq:phi0_bdry},\ref{eq:phi1_bdry}) is fixed and independent of
the configuration.




\subsection{Application to the dilute loop model}
\label{sec:bound-dil}
We consider the dilute loop model on the light-cone lattice, with as in
the dense case possible defects located next to each other on the bottom row.
All the observables that we consider force a line to go from the insertion
point to the boundary defect, so that a typical
configuration is as depicted below.

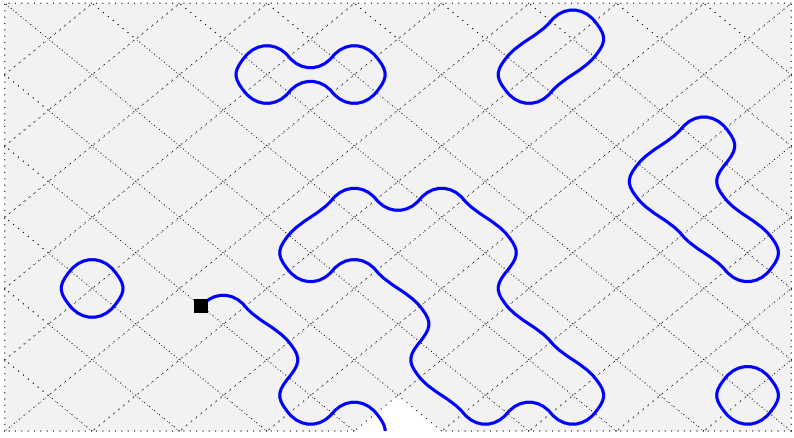
\begin{center}
\slow{
\begin{tikzpicture} [scale=0.75,rotate=\lcrot,distort]
\newcount\u\newcount\v
\pgfmathsetseed{45}
\foreach\x in {-7,...,7}
\foreach\y in {-7,...,7}
{\pgfmathrandominteger{\rand}{0}{1}
\pgfmathsetcount{\u}{\x+\y}
\pgfmathsetcount{\v}{\x-\y}
\ifnum\u<6
  \ifnum\u>-6
     \ifnum\v<9
        \ifnum\v>-9
           \plaqz(\x,\y)
        \else
        \fi
     \else
     \fi
   \else
   \fi
\else
\fi
\ifnum\u=6
  \bplaqnz(\x,\y)
\else
  \ifnum\u=-6
     \ifnum\y=-3
     \else
        \bplaqsz(\x,\y)
     \fi
  \else
  \fi
\fi
\ifnum\v=9
   \bplaqez(\x,\y)
\else
  \ifnum\v=-9
     \bplaqwz(\x,\y)
  \else
  \fi
\fi
}
\begin{scope}[shift={(-3,-3)}]
\clip (-4.5,4.5) -- (4.5,-4.5) -- (4.5,-0.5) -- (-0.5,4.5) -- cycle;
\draw[edge] (-0.5,0) \start\east\go\north\go\west\go\north\go\east\go\north\go\north\go\west node[oper] {};
\end{scope}
\draw[edge] (2,2.5) \start\north\go\west\go\north\go\west\go\south\go\east\go\south\go\east\go\north;
\draw[edge] (5,-2.5) \start\north\go\west\go\west\go\south\go\south\go\south\go\east\go\north\go\north\go\east\go\north;
\draw[edge] (0,0.5) \start\south\go\east\go\south\go\south\go\west\go\south\go\south\go\south\go\west\go\north\go\west\go\north\go\north\go\east\go\north\go\north\go\west\go\north\go\east\go\east\go\south;
\draw[edge] (-5,2.5) \start\south\go\east\go\north\go\west\go\south;
\draw[edge] (3,0.5) \start\south\go\east\go\east\go\north\go\west\go\west\go\south;
\draw[edge] (1,-6.5) \start\south\go\east\go\north\go\west\go\south;
\end{tikzpicture}
}
\end{center}


\subsubsection{Loop observables in the bulk}
In contrast with the dense case, the $K$-matrix~\eqref{eq:K-19V} does not
introduce any orientation-dependent phase factor, but simply a relative
weight for contacts with the boundary. Therefore, the analysis of
observables $E_0$ and $E_1$ 
in the bulk is unchanged, and we do not repeat it here.

A natural question is whether one can use the adjoint action to build 
new currents. 
We consider the element $P \in \Uq{A^{(2)}_2}$ given by 
$P =   \ad{E_1}{E_0}$.
Explicitly,
\begin{align*}
P
= E_1 E_0 - T_1 E_0 T_1^{-1} E_1
  = E_1 E_0 -q^{-4} E_0 E_1 \,,
\end{align*}
and hence we can use the treatment of \secref{sec:adj-lc}
to express the corresponding lattice observable $p$.
From~\eqref{eq:adjoint-defect}, we have
\[
\aver{p(x,t)} = 
\aver{A_{E_1}\left[e_0(x,t)\right]} =
\aver{\left[ \wh{e}_1(x_d,0) + \wh{e}_1(x_d+1,0)\right] e_0(x,t)} \,.
\]
Since $E_1$ can only flip a state $\downarrow$ to a state $\uparrow$, the non-zero terms
correspond to a pair of defects $(\downarrow,0)$ or $(0,\downarrow)$.
Summing over these two possibilities, we get:
$\aver{p(x,t)} = -q^{-4} (z^{2\ell}+z^{-2\ell})
\aver{ e_0(x,t)}$.
Hence, the operator $P$ does not lead to a new holomorphic observable; it simply
produces $\phi_0$, up to a multiplicative constant.

Equivalently, let us define 
\begin{equation*}
  \xi(x,t) := z^{\ell-1} 
  \begin{cases}
    p(x,t), & x+t=1\pmod 2 \\ 
    e^{i\alpha} p(x,t), & x+t=0\pmod 2\,, 
  \end{cases}
  \qquad
  \phi_0(x,t)
:=
z^{\ell-1}
\begin{cases}
    e_0(x,t), & x+t=1\pmod 2 \\ 
    e^{i\alpha} e_0(x,t), & x+t=0\pmod 2\,.
  \end{cases}
\end{equation*}
Then
\begin{equation}\label{eq:pvse0}
\aver{\xi(x,t)}=c\,\aver{\phi_0(x,t)},
\qquad
\text{where}\ \ c=-q^{-4} (z^{2\ell}+z^{-2\ell}).
\end{equation}
Since it is only possible to relate $\xi$ and $\phi_0$ as expectation values $\aver{\cdots}$, the constant $c$ is dependent on the choice of boundary conditions; it would differ, had we chosen alternative boundaries to the ones shown in the figure above.

\subsubsection{Loop observables at the boundary}
We are now in a position to construct an observable which satisfies discrete 
holomorphicity at the left boundary. We consider the operator
$Q = P + iq^{-1} \bar E_0$ (with $P$ as in the previous section),
which commutes
with the $K$-matrix in the sense of \eqref{eq:Kcomm}, \cf \S \ref{sssec:bound-alg-2}.

>From \eqref{eq:leftconserv} with $j_a=p+iq^{-1}\bar e_0$ we obtain 
\begin{equation*}
p(1,t) + iq^{-1} \bar e_0(1,t)
=
p(1,t+1) + iq^{-1} \bar e_0(1,t+1)\,,
\end{equation*}
which can be translated into an equation for the observables $\xi$ and $\bar\phi_0$:
\begin{equation*}
z^{1-\ell} \xi(1,t)
  + 
iq^{-1}z^{\ell-1} \bar\phi_0(1,t)
 =
e^{-i\alpha} z^{1-\ell} \xi(1,t+1)  
+ 
e^{i\alpha} iq^{-1} z^{\ell-1} \bar\phi_0(1,t+1)
 \,.
 \end{equation*}
Taking the real part of this equation, we obtain
\begin{equation*}
  \mathrm{Re} \left[\psi(1,t) + e^{i(\pi-\alpha)} \psi(1,t+1) \right] = 0 \,,
\end{equation*}
where $\psi=z^{1-\ell}(\xi-iq\phi_0)$.
Now taking into account \eqref{eq:pvse0},
we find that in the equation above one can replace
$\psi$ with $e^{i\lambda}\phi_0$ ($\lambda$ being some phase, dependent on our choice of boundary conditions, which we do not write explicitly).


\section{The continuum limit}
\label{sec:continuum}

\subsection{Dense loops}
In this section, we identify the operators corresponding to the lattice
holomorphic observables, as holomorphic currents in the
CFT describing the continuum limit.

\subsubsection{Scaling theory}
Let us first briefly review the Coulomb gas construction~\cite{Nienhuis84,DFSZ87}.
We define a height function $\Phi(x+1/2,t+1/2)$ on the dual lattice, such that
the oriented loops are the contour lines of $\Phi$, and with
the convention that the values of $\Phi$ across a
contour line differ by $\pi$. In the continuum limit, the 
coarse-grained height function
is then subject to a Gaussian distribution:
\begin{equation} \label{eq:boson}
  S[\Phi] = \frac{g}{4\pi} \int (\nabla\Phi)^2 dx dt \,,
  \qquad \text{where} \qquad
  g = 1-2\nu \,.
\end{equation}
Note that the coupling constant spans the interval $0<g<1$.
In what follows, the model is defined on a cylinder of circumference $L$,
and the axis of the cylinder is in the time direction.
Because of the definition of $\Phi$ by local increments, $\Phi$ can be
discontinuous along the circumference:
\[
\Phi(L+x+1/2,t+1/2) - \Phi(x+1/2,t+1/2) = \pi m \,,
\qquad m \in \mathbb{Z} \,.
\]
Thus, the height function should be considered as living on a circle:
\[
 \Phi \equiv \Phi +\pi \,.
\]
Note that, for this dense loop model, when $L$ is even (resp. odd),
only even integer (resp.\ odd integer) values of $m$ are reached.
Also, the local Boltzmann weights associated to loop turns ensure that
the closed loops get the correct weight $\tau=2\cos(2\pi\nu)$, except for the
non-contractible loops: these loops have a vanishing total winding, and thus get a weight
$\widetilde{\tau}=2$. To restore the correct weight, one introduces a seam in the time direction,
such that every right (resp.\ left) arrow crossing the seam gets a weight $e^{i\pi\alpha}$ (resp.\ $e^{-i\pi\alpha}$).
The weight of non-contractible loops becomes $\widetilde{\tau}=2\cos\pi\alpha$, and one sets $\alpha:=2\nu$
to get $\widetilde{\tau}=\tau$.

\subsubsection{Operator content}
To recover the full-plane geometry, we use the complex coordinates
\[
z = e^{2\pi(t+ix)/L} \,, \qquad \bar{z} = e^{2\pi(t-ix)/L} \,.
\]
In this setting, the seam described above goes from the origin to infinity, and
amounts to introducing a pair of vertex operators
$e^{i\alpha \Phi(\infty)} e^{-i\alpha \Phi(0)}$.
More generally, if we decompose the height field as $\Phi(z,\bar{z})= \varphi(z)+\bar{\varphi}(\bar{z})$,
the primary operators are of the form $\mathcal{O}_{\mu,\bar{\mu}} = e^{i(\mu\varphi+\bar{\mu}\bar{\varphi})}$, which we write as
\[
\mathcal{O}_{\mu,\bar{\mu}} = e^{i(\mu+\bar{\mu})(\varphi+\bar{\varphi})/2}
\times e^{i(\mu-\bar{\mu})(\varphi-\bar{\varphi})/2} \,.
\]
The first factor is only well-defined if $n:=(\mu+\bar{\mu})/4$ is an integer,
which is called the electric charge $n \in \mathbb{Z}$.
The average value of the
second factor is $e^{i \Phi_{\rm cl}}$, where
$\Phi_{\rm cl} = i(\mu-\bar{\mu})(\log z - \log \bar{z})/(4g)$.
The discontinuity of $\Phi_{\rm cl}$ around the origin is $\delta\Phi_{\rm cl}=\pi\times (\mu-\bar{\mu})/g$, and
hence the number $m:=(\mu-\bar{\mu})/g$ must be an integer, and is called the magnetic charge
$m \in \mathbb{Z}$.
Since the conformal weight of the chiral vertex operator $e^{i\mu\varphi}$ is $h_\mu = \mu (\mu+2\alpha)/(4g)$,
we get for $\mathcal{O}_{\mu,\bar{\mu}}$:
\[
h = \frac{(2n+\alpha + mg/2)^2 - \alpha^2}{4g} \,,
\qquad
\bar{h} = \frac{(2n+\alpha - mg/2)^2 - \alpha^2}{4g} \,,
\]
In this context, $\alpha$ appears as a background electric charge.
Part of the above spectrum fits in the Kac table for minimal models:
\[
h_{r,s} = \frac{(r-gs)^2 - (1-g)^2}{4g} \,,
\]
with integer $r,s$.

\subsubsection{Scaling limit of the lattice observables}\label{sec:lattobs}
We will now show that the discrete (anti-)holomorphic observables discussed in this paper scale
to operators of the form $\mathcal{O}_{\mu,\bar{\mu}}$.
Let us first consider the two-point function associated to $E_0$:
\[
\aver{\phi_0(z,\bar{z}) \phi_0^*(w,\bar{w})} :=
\frac{1}{Z} \sum_{C|\gamma_1,\gamma_2:w\to z} e^{i(4\nu-1)\theta} \ W(C) \,,
\]
where the sum is on loop configurations with two open oriented paths $\gamma_1,\gamma_2$ going from $w$ to $z$, and $\theta$ is the winding
angle of each path. In terms of the height model, this correlation function includes magnetic defects of charges $-2$ and $+2$
at $z$ and $w$, respectively. Moreover, the winding angle is given by $\theta=\Phi(z,\bar{z})-\Phi(w,\bar{w})$, and
hence the factor $e^{i(4\nu-1)\theta}$ corresponds to an electric operator of charge $2n+\alpha=4\nu-1$ at $z$, and opposite charge at $w$. These charges yield the values $\mu = -2g$ and $\bar{\mu}=0$, and thus we identify:
\[
\phi_0 = \phi_0(z) = e^{-2ig \varphi(z)} \,,
\]
with conformal dimensions $h=2g-1=h_{13}$ and $\bar{h}=0$. By reversing the arrows, we obtain
\[
\bar\phi_0(\bar{z}) = e^{-2ig \bar{\varphi}(\bar{z})} \,.
\]

Similarly, the two-point function associated to $E_1$ is:
\[
\aver{\phi_1(z,\bar{z}) \phi_1^*(w,\bar{w})} :=
\frac{1}{Z} \sum_{C|\gamma_1,\gamma_2:w\to z} e^{i\theta} \ W(C) \,,
\]
which has charges $m=2$ and $2n+\alpha = 1$, and hence we get
\[
\phi_1(z) = e^{2ig \varphi(z)} \,,
\qquad
\bar\phi_1(\bar z) = e^{2ig \bar{\varphi}(\bar{z})} \,.
\]
The holomorphic current $\phi_1$ has conformal dimensions $h=1$ and $\bar{h}=0$: this is the
``screening operator''.

We now turn to the lattice operator associated to the diagonal generators $H_i$.
Since $H_i \propto \sigma^z$, it simply measures the local orientation of loops.
The increment of $\Phi$ across an up (resp.\ down) arrow
is $\pi$ (resp.\ $-\pi$), and thus one has $a \partial_x \Phi = \pi h^{(\tsup)}$, and likewise $a \partial_t \Phi = -\pi h^{(\xsup)}$,
where $a$ is the lattice mesh size. Using the complex coordinates $w=t+ix$ and $\bar w=t-ix$,
we identify the chiral currents:
\[
h^{(\xsup)}+ih^{(\tsup)} \propto \partial_w \varphi \,,
\qquad
h^{(\xsup)}-ih^{(\tsup)} \propto \partial_{\bar w} \bar\varphi \,,
\]
The conservation law found in \secref{sec:localobs} corresponds in the continuum
limit to the 
conservation of non-chiral current $d^\ast \Phi=d^\ast(\varphi+\bar\varphi)$, 
or in components, $\epsilon^{\mu\nu}\partial_\nu \Phi$, which ensures local 
well-definedness of $\Phi$.

The above results can be summarised in the following diagram:
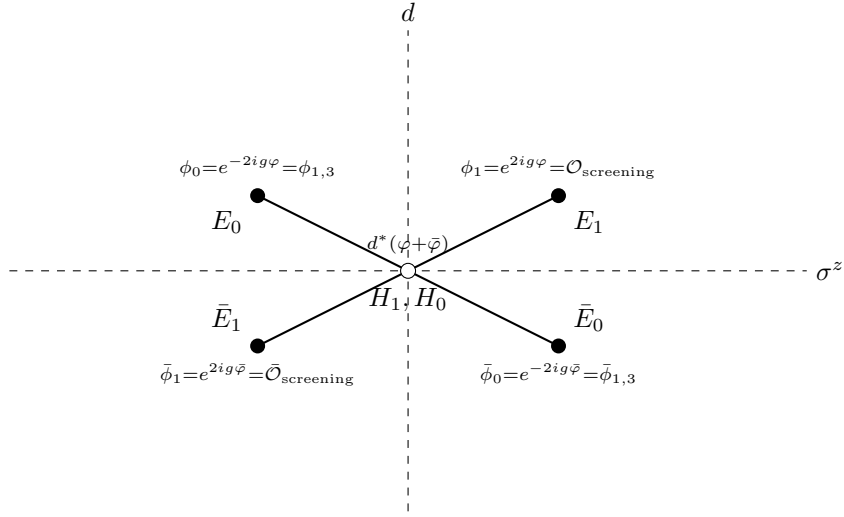
\begin{center}
  \begin{tikzpicture}
    \draw[dashed] (0,-3.2) -- (0,3.2) node[above] {$d$};
    \draw[dashed] (-5.3,0) -- (5.3,0) node[right] {$\sigma^z$};
    \draw[thick] (-2,-1) -- (2,1);
    \draw[thick] (-2,1) -- (2,-1);
    \node[circle,draw,fill=white,inner sep=2pt,label={below:$H_1,H_0$},label={above:$\ss d^\ast(\varphi+\bar\varphi)$}] 
    at (0,0) {};
    \node[circle,fill,inner sep=2pt,label={below right:$E_1$},label={above:$\ss
        \phi_1=e^{2ig\varphi}=\mathcal{O}_{\rm screening}$}]
    at (2,1) {};
    \node[circle,fill,inner sep=2pt,label={above left:$\bar E_1$},label={below:$\ss
        \bar\phi_1=e^{2ig\bar\varphi}=\bar{\mathcal{O}}_{\rm screening}$}]
    at (-2,-1) {};
    \node[circle,fill,inner sep=2pt,label={below left:$E_0$},label={above:$\ss
        \phi_0=e^{-2ig\varphi}=\phi_{1,3}$}] 
    at (-2,1) {};
    \node[circle,fill,inner sep=2pt,label={above right:$\bar E_0$},label={below:$\ss
        \bar\phi_0=e^{-2ig\bar\varphi}=\bar\phi_{1,3}$}] 
    at (2,-1) {};
  \end{tikzpicture}
\end{center}
In this figure, the horizontal axis is the $U(1)$ charge $\sigma^z$, and the vertical
axis is the gradation $d$ in the evaluation representation of $A_1^{(1)}$.
We note the similarity with the discussion
of nonlocal charges in the (ultra-violet limit of the) 
sine--Gordon model in~\cite{BernardLeclair91}.
A notable difference is the choice of gradation, which has different origins
in the two situations. 

\subsection{Dilute loops}
The mapping to a compactified free boson CFT~\eqref{eq:boson} also holds in the dilute case, up
to minor adaptations. We keep the convention that the height function~$\Phi$ has jumps of $\pm\pi$
across an oriented loop,
and, in the scaling theory, one should set $\Phi \equiv \Phi+\pi$. So we can keep the same notations as in the
previous section, except 
that $\nu$ is now chosen in the interval $[-1/2,0]$, and we have $1<g<2$.

The two-point function for $\phi_0$ is:
\[
\aver{\phi_0(z,\bar{z}) \phi_0^*(w,\bar{w})} =
\frac{1}{Z} \sum_{C|\gamma:w\to z} e^{i(3\nu/2-1/4)\theta} \ W(C) \,,
\]
where the sum is on loop configurations with one open oriented path $\gamma$ going from $w$ to 
$z$, and $\theta$ is the winding angle of $\gamma$. Since this is a one-leg defect and
$\theta = 2[\Phi(z,\bar z)-\Phi(w,\bar w)]$, the corresponding charges are $m=-1$ and $2n+\alpha=
3\nu-1/2$. The ``flux observables'' $\phi_1$ and $\bar \phi_1$ are the same as in
the dense model. Thus we obtain
\begin{align*}
  \phi_0(z) = e^{-ig \varphi(z)} \,,
  &\qquad
  \bar\phi_0(\bar z) = e^{-ig \bar\varphi(\bar z)} \,, \\
  \phi_1(z) = e^{2ig \varphi(z)} \,,
  &\qquad
  \bar\phi_1(\bar z) = e^{2ig \bar\varphi(\bar z)} \,,
\end{align*}
and the conformal dimension for $\phi_0$ is $h=(3g-2)/4=h_{12}$, whereas
for $\phi_1$ it is $h=1$.
Finally, the diagonal operators
$h^{(x,t)}$ relate to $\partial_w \varphi$ and $\partial_{\bar w} \bar\varphi$.

\section{Conclusions}\label{sec:conclusion}
In this paper, we have described a general procedure to obtain discretely
holomorphic observables out of nonlocal currents in quantum integrable 
lattice models. We have shown in several examples how these
observables are naturally expressed in terms of loop models. We have
identified them in the continuum limit, connecting to Conformal Field Theory.

It should be noted that in CFT the conserved currents
always come in pairs: a current $j^\mu$ and its dual current $\tilde\jmath^\mu=\epsilon^{\mu}_\nu j^\nu$. Only the two conservation laws combined
imply separation of chiralities, and therefore
existence of holomorphic observables. Here we only base our analysis on a single
conservation law for each observable, hence a ``weak'' discrete holomorphicity
condition -- the dual equation is missing. This absence can be traced
to the step in which we identify the two components (say, time and space) of
our current as a single function corresponding to the observable. This step
would require additional justification in order to proceed with a rigorous
proof of the conformal limit.\footnote{In fact, such an identification is not
possible for the nonchiral observable associated to $H_1$, \cf \secref{sec:lattobs}.} The fact that in all cases,
our would-be chiral observables have, in the loop language, a unifying
definition on both vertical and horizontal edges is certainly a strong
indication that such an identification is correct.

This work opens the way to further study and interpretation of discrete
holomorphic observables, in particular in the case of more general
boundary conditions (as recently studied in~\cite{deGierLR12}).
Also, the application of this approach to off-critical models (see
the treatment of the Ising model in~\cite{RivaC06}) needs to be
developed.

\bibliographystyle{amsplainhyper}
\bibliography{biblio}

\providecommand{\bysame}{\leavevmode\hbox to3em{\hrulefill}\thinspace}
\begin{thebibliography}{10}

\bibitem{AlamBatchelor12}
I.~{Alam} and M.~{Batchelor}, \emph{{Integrability as a consequence of discrete
  holomorphicity: the {$\mathbb{Z}_N$} model}}, 2012, preprint,
  \href{http://arxiv.org/abs/1207.3883}{\path{arXiv:1207.3883}}.

\bibitem{BFKZ96}
M.~T. Batchelor, V.~Fridkin, A.~Kuniba, and Y.~K. Zhou, \emph{{Solutions of the
  reflection equation for face and vertex models associated with $A_n^{(1)},
  B_n^{(1)}, C_n^{(1)}, D_n^{(1)}$ and $A_n^{(2)}$}}, Physics Letters B
  \textbf{376} (1996), no.~4, 266 -- 274.

\bibitem{Bax82}
R.~J. Baxter, \emph{Exactly solved models in statistical mechanics}, second
  ed., Dover, USA, 2007, Reprint of the 1982 original (Academic Press, London).

\bibitem{BeatondGG11}
N.~{Beaton}, J.~{de Gier}, and A.~{Guttmann}, \emph{{The critical fugacity for
  surface adsorption of {SAW} on the honeycomb lattice is $1+\sqrt{2}$}},
  September 2011, preprint,
  \href{http://arxiv.org/abs/1109.0358}{\path{arXiv:1109.0358}}.

\bibitem{BeatonGJ12}
N.~Beaton, A.~Guttmann, and I.~Jensen, \emph{{T}wo-dimensional self-avoiding
  walks and polymer adsorption: critical fugacity estimates}, J. Phys. A
  \textbf{45} (2012), no.~5, 055208,
  \url{http://stacks.iop.org/1751-8121/45/i=5/a=055208}.

\bibitem{BernardFelder91}
D.~Bernard and G.~Felder, \emph{Quantum group symmetries in two-dimensional
  lattice quantum field theory}, Nucl. Phys. B \textbf{365} (1991), no.~1, 98
  -- 120, \href{http://dx.doi.org/10.1016/0550-3213(91)90608-Z}{\path{doi}}.

\bibitem{BernardLeclair91}
D.~Bernard and A.~Leclair, \emph{Quantum group symmetries and non-local
  currents in 2d {QFT}}, Commun. Math. Phys. \textbf{142} (1991), 99--138
  (English), \href{http://dx.doi.org/10.1007/BF02099173}{\path{doi}}.

\bibitem{chpr94}
V.~Chari and A.~Pressley, \emph{A guide to quantum groups}, Cambridge, 1994.

\bibitem{ChelkakHI12}
D.~{Chelkak}, C.~{Hongler}, and K.~{Izyurov}, \emph{Conformal invariance of
  spin correlations in the planar {I}sing model}, February 2012, preprint,
  \href{http://arxiv.org/abs/1202.2838}{\path{arXiv:1202.2838}}.

\bibitem{ChelkakS09}
D.~{Chelkak} and S.~{Smirnov}, \emph{{Universality in the 2D {I}sing model and
  conformal invariance of fermionic observables}}, October 2009, preprint,
  \href{http://arxiv.org/abs/0910.2045}{\path{arXiv:0910.2045}}.

\bibitem{deGierLR12}
J.~de~Gier, A.~Lee, and J.~Rasmussen, \emph{Discrete holomorphicity and
  integrability in loop models with open boundaries}, October 2012, preprint,
  \href{http://arxiv.org/abs/1210.5036}{\path{arXiv:1210.5036}}.

\bibitem{DFSZ87}
P.~Di~Francesco, H.~Saleur, and J.-B. Zuber, \emph{{R}elations between the
  {C}oulomb gas picture and conformal invariance of two-dimensional critical
  models}, J. Stat. Phys. \textbf{49} (1987), 57--79,
  \href{http://dx.doi.org/10.1007/BF01009954}{\path{doi}}.

\bibitem{DotsenkoP88}
Vl. Dotsenko and A.~Polyakov, \emph{Fermion representations for the 2d and 3d
  {I}sing models}, Adv. Stud. Pure Math. \textbf{16} (1988), 171--203,
  Conformal field theory and solvable lattice models, Ed. by M. Jimbo, T. Miwa,
  A. Tsuchiya, Academic Press, 426 p. (1988). ISBN 0-12-385340-0.

\bibitem{DuminilS10}
H.~{Duminil-Copin} and S.~{Smirnov}, \emph{{The connective constant of the
  honeycomb lattice equals $\sqrt{2+\sqrt{2}}$}}, July 2010, preprint,
  \href{http://arxiv.org/abs/1007.0575}{\path{arXiv:1007.0575}}.

\bibitem{FendleyTalk12}
P.~Fendley, \emph{Discrete holomorphicity from topology}, talk given at the
  conference ``Conformal Invariance, Discrete Holomorphicity and
  Integrability", Helsinki, June 2012.

\bibitem{HonglerK11}
C.~Hongler and K.~{Kyt\"ol\"a}, \emph{{I}sing interfaces and free boundary
  conditions}, August 2011, preprint,
  \href{http://arxiv.org/abs/1108.0643}{\path{arXiv:1108.0643}}.

\bibitem{HonglerKZ12}
C.~Hongler, K.~Kyt{\"o}l{\"a}, and A.~Zahabi, \emph{Discrete holomorphicity and
  {I}sing model operator formalism}, 2012, preprint,
  \href{http://arxiv.org/abs/1211.7299}{\path{arXiv:1211.7299}}.

\bibitem{Ikhlef12}
Y.~Ikhlef, \emph{Discretely holomorphic parafermions and integrable boundary
  conditions}, J. Phys. A \textbf{45} (2012), no.~26, 265001,
  \url{http://stacks.iop.org/1751-8121/45/i=26/a=265001}.

\bibitem{IkhlefC09}
Y.~Ikhlef and J.~Cardy, \emph{{D}iscretely holomorphic parafermions and
  integrable loop models}, J. Phys. A \textbf{42} (2009), no.~10, 102001.

\bibitem{Jim85}
M.~Jimbo, \emph{{A q-analogue of $U(gl(N+1))$, Hecke algebra, and the
  Yang-Baxter equation }}, Lett. Math. Phys. \textbf{10} (1985), 63.

\bibitem{Jimbo86}
M.~Jimbo, \emph{Quantum {$R$}-matrix for the generalized {T}oda system},
  Commun. Math. Phys. \textbf{102} (1986), 537--547,
  \href{http://dx.doi.org/10.1007/BF01221646}{\path{doi}}.

\bibitem{NepMez98}
L.~Mezincescu and R.~I. Nepomechie, \emph{Fractional-spin integrals of motion
  for the boundary sine-{G}ordon model at the free fermion point}, Internat. J.
  Modern Phys. A \textbf{13} (1998), no.~16, 2747--2764.

\bibitem{Nep02}
R.~I. Nepomechie, \emph{Boundary quantum group generators of type {$A$}}, Lett.
  Math. Phys. \textbf{62} (2002), no.~2, 83--89.

\bibitem{Nienhuis84}
B.~Nienhuis, \emph{{C}ritical behavior of two-dimensional spin models and
  charge asymmetry in the {C}oulomb gas}, J. Stat. Phys. \textbf{34} (1984),
  731--761, \href{http://dx.doi.org/10.1007/BF01009437}{\path{doi}}.

\bibitem{RajC07}
M.~Rajabpour and J.~Cardy, \emph{Discretely holomorphic parafermions in lattice
  {$\mathbb{Z}_N$} models}, J. Phys. A \textbf{40} (2007), no.~49, 14703,
  \url{http://stacks.iop.org/1751-8121/40/i=49/a=006}.

\bibitem{RivaC06}
V.~Riva and J.~Cardy, \emph{{H}olomorphic parafermions in the {P}otts model and
  stochastic {L}oewner evolution}, J. Stat. Mech. \textbf{2006} (2006), no.~12,
  P12001.

\bibitem{Skl88}
E.~Sklyanin, \emph{{B}oundary conditions for integrable quantum systems}, J.
  Phys. A \textbf{21} (1988), no.~10, 2375--2389.

\bibitem{Smir07}
S.~Smirnov, \emph{{C}onformal invariance in random cluster models {I}.
  {H}olomorphic fermions in the {I}sing model}, Ann. Math. \textbf{172} (2010),
  no.~2, 1435--1467,
  \href{http://dx.doi.org/10.4007/annals.2010.172.1435}{\path{doi}}.

\end{thebibliography}

\end{document}